\documentclass[12pt]{article}
\usepackage{amsmath}
\usepackage{amsfonts}
\usepackage{amsthm}
\usepackage{mathrsfs}
\usepackage{graphicx,psfrag,epsf}
\usepackage{enumerate}
\usepackage{natbib}
\usepackage{tabularx, ragged2e}
\usepackage{booktabs}
\usepackage{color}
\usepackage{url} % not crucial - just used below for the URL
\usepackage{apxproof}
\usepackage{enumitem}
\usepackage[toc,title]{appendix}

\newtheorem{theorem}{Theorem}
\newtheorem{corollary}{Corollary}[theorem]
\newtheorem{lemma}{Lemma}
\newtheorem{proposition}{Proposition}

\theoremstyle{definition}
\newtheorem{example}{Example}

\newcommand{\blind}{1}
\newcolumntype{Y}{>{\centering\arraybackslash}X}

\begin{document}

	\def\spacingset#1{\renewcommand{\baselinestretch}%
		{#1}\small\normalsize} \spacingset{1}

	%%%%%%%%%%%%%%%%%%%%%%%%%%%%%%%%%%%%%%%%%%%%%%%%%%%%%%%%%%%%%%%%%%%%%%%%%%%%%%
	
	\if1\blind
	{
		\title{\bf Conditionally Affinely Invariant Rerandomization and its Admissible Complete Class}
		\author{Zhen Zhong\\ Faculty of Business and Economics, The University of Hong Kong \\
			Donald B. Rubin\\
			Department of Statistical Science, Temple University}
		\maketitle
	} \fi
	
	\if0\blind
	{
		\bigskip
		\bigskip
		\bigskip
		\begin{center}
			{\LARGE\bf Title}
		\end{center}
		\medskip
	} \fi
	
	\bigskip
	\spacingset{1.5}
	\begin{abstract}
		Rerandomization utilizes modern computing ability to improve covariate balance while adhering to the randomization principle originally advocated by RA Fisher. Affinely invariant rerandomization has the ``Equal Percent Variance Reducing'' (EPVR) property. When dealing with covariates of varying importance and/or mixed types, the conditionally EPVR property is often more desired. We discuss a general class of conditionally affinely invariant rerandomization methods and obtain their conditionally EPVR property. In addition, we set up a decision-theoretical framework to evaluate balance criteria for rerandomization. Popular rerandomization methods, such as the covariate balance table check, are found to be inadmissible. We suggest an admissible complete class of conditionally affinely invariant balance criteria, which can be applied to experimental designs involving tiers of covariates, stratification, and multiple treatment arms.
	\end{abstract}
	
	\noindent%
	{\it Keywords: Randomized experiments; ellipsoidal distribution; statistical decision theory}
	\vfill
	
	\newpage
	\spacingset{2} % DON'T change the spacing!
	\section{Introduction}
	\label{sec:intro}
	Randomized controlled trials (RCTs) are the ``gold standard" for studying causal relationships, yet can be expensive to implement. \cite{tukey1993tightening} describes the ``platinum standard" for RCTs, where probability statements, such as statistical significance, depend only on exactly how the trial was conducted. Extensive covariates are now commonly available at the design stages of RCTs. According to a survey by \cite{bruhn2009pursuit}, researchers tend to rerandomize when imbalance in covariates across treatments occurs but the exact rerandomization procedures are generally not described in detail. Any subsequent analysis that ignores the design stage does not adhere to Tukey's ``platinum standard".
	
	\citeauthor{morgan2012rerandomization} (MR, 2012) formalize rerandomization by specifying a balance criterion before treatment assignment, accepting a randomized assignment only after the criterion is met, and then analyzing results using randomization-based methods appropriate to that design. The results of rerandomization procedures are transparent and ensure the validity of Fisher's randomization test \citep{fisher1935design}. MR focus on rerandomization methods with affinely invariant balance criteria. Using symmetry arguments, rerandomization based on the Mahalanobis distance is shown to be ``Equal Percent Variance Reducing" (EPVR).
	
	Affine invariance can be relaxed to apply to tiers of covariates \citep[e.g.]{morgan2015rerandomization}, special discrete covariates \citep[e.g.]{wang2021rerandomization,johansson2022rerandomization}, and factorial designs \cite[e.g.][]{branson2016improving,li2020rerandomization}. A general theory for such designs is our first major topic. Parallel to \cite{rubin1992affinely}, we make extensions to the affinely invariant balance criteria and obtain the conditionally EPVR property with respect to the limiting covariate distribution assumed to be conditionally ellipsoidally symmetric. This assumption can be satisfied by applying the finite population central limit theorem in each scenario.
	
	In contrast, other researchers seek optimally balanced designs (e.g., \citealp{bertsimas2015power}, \citealp{kallus2018optimal}). However, the sets of optimally balanced assignments are too restrictive to allow valid randomization-based inference \citep{johansson2021optimal} and can cause imbalances of unobserved covariates \citep{harshaw2024balancing}. Therefore, it's necessary to keep a minimal acceptance probability so that meaningful randomization-based inference can be drawn. Inspired by the statistical decision theory \citep{wald1949statistical}, we give formal definitions of admissibility and complete class for stochastic criteria, and characterize them with sufficient and necessary conditions. ``Unified" and ``intersection" methods, as defined in Section \ref{sec:decision_theory}, are two common ways of building balance criteria. Despite their popularity, many ``intersection'' balance criteria, including the covariate balance table check \citep[see e.g.][]{zhao2024no}, are found to be inadmissible. We suggest an admissible complete class of conditionally affinely invariant balance criteria, which is found to dominate existing methods for multiple balancing targets.
	
	We organize the rest materials as follows. Section \ref{sec:review} reviews affinely invariant rerandomization methods. Section \ref{sec:main} motivates the idea conditionally affinely invariant rerandomization methods and studies their basic properties. Section \ref{sec:decision_theory} characterizes admissible/inadmissible balance criteria and compares ``unified" methods with ``intersection" methods. Section \ref{sec:rewm} studies the admissible complete class of conditionally affinely invariant balance criteria. Section \ref{sec:example} illustrates our theory using a real dataset from \cite{blattman2017reducing}. Section \ref{sec:d&c} makes further discussion and concludes.
	
	\section{A Review of Affinely Invariant Rerandomization}
	\label{sec:review}
	Let's firstly consider a randomized experiment with two treatment levels over $n$ units. Let $\boldsymbol{W}$ denote the treatment allocation vector of dimension $n$, whose $i$-th value $W_i$ equals $1$ if unit $i$ ($i=1,2,\ldots,n$) is treated and equals $0$ if not. Under complete randomization, a treatment allocation is randomly drawn from $\mathcal{W}=\{\boldsymbol{W}\in\{0,1\}^n|\sum{}W_i=n_{\text{1}}\}$, where $n_{\text{1}}$ is the treatment group size, and $n_{\text{0}}=n-n_{\text{1}}$ is the control group size. 
	
	When experimental units are enrolled in a study, they typically include their baseline covariates such as age and gender, that could potentially affect experimental outcomes. Let  $\boldsymbol{x}$ be an $n\times{}k$ real-valued matrix in which the $i$'th row $\boldsymbol{x}_i$ is the covariate vector for unit $i$. Under complete randomization, a random draw of treatment allocation from $\mathcal{W}$ can be highly correlated with certain linear combination of covariates. It's not a good idea to conduct the experiment with such an ``unlucky" randomization because covariates are not balanced between treatment and control groups.
	
	Rerandomization rejects a random draw from $\mathcal{W}$ if some pre-determined balance criterion is not satisfied. The criterion can be formulated as an indicator function $\varphi(\boldsymbol{x}, \boldsymbol{W})$, where a decision is made to accept $\boldsymbol{W}$ if $\varphi=1$ and if $\varphi=0$ to reject $\boldsymbol{W}$ and rerandomize. A balance criterion is said to be affinely invariant if it satisfies 
	\begin{equation*}
		\varphi\left(\mathcal{A}(\boldsymbol{x}), \boldsymbol{W}\right)=\varphi\left(\boldsymbol{x}, \boldsymbol{W}\right)
	\end{equation*}
	for any affine transformation $\mathcal{A}$ on covariates defined as $\mathcal{A}(\boldsymbol{x})=\boldsymbol{x}\boldsymbol{A}+\boldsymbol{b}$.
	\begin{example}[ReM]
		\label{rem}
		MR suggest rerandomization based on the squared Mahalanobis distance between treatment and control groups (ReM):
		\begin{equation*}
			D\equiv \widehat{\boldsymbol{\tau}}_{\boldsymbol{x}}\operatorname{cov}\left(\widehat{\boldsymbol{\tau}}_{\boldsymbol{x}}\right)^{-1}\widehat{\boldsymbol{\tau}}_{\boldsymbol{x}}^{\top},
		\end{equation*}
		where $\widehat{\boldsymbol{\tau}}_{\boldsymbol{x}}=\overline{\boldsymbol{x}}_{\text{1}}-\overline{\boldsymbol{x}}_{\text{0}}$ is the difference in treatment and control group means of covariates, and $\operatorname{cov}\left(\widehat{\boldsymbol{\tau}}_{\boldsymbol{x}}\right)$ is the sampling covariance of $\widehat{\boldsymbol{\tau}}_{\boldsymbol{x}}$ under complete randomization. A randomization is rejected if $D>a$ and accepted otherwise, where $a>0$ is a pre-specified threshold. This balance criterion is affine invariant because $D$ remains invariant under any affine transformation $\mathcal{A}$ on covariates.
		
		A measure of covariate balance proposed by MR is the percentage reduction in variance of the difference in means for each covariate. MR show that, using ReM, the reduction in variance for each covariate, and for any linear combination of covariates is the same, namely EPVR. The conclusion builds on a finite population central limit theorem by \cite{hajek1960limiting}, which in fact holds with respect to the limiting distribution of $\sqrt{n}\widehat{\boldsymbol{\tau}}_{\boldsymbol{x}}$. Let $\mathbb{E}_{\text{lim}}$, $\operatorname{var}_{\text{lim}}$ and  $\operatorname{cov}_{\text{lim}}$ denote the mean, variance and covariance with respect to the limiting distribution, respectively. Let $\sqrt{n}\widehat{\boldsymbol{\tau}}_{\boldsymbol{x}}\boldsymbol{\gamma}$, $\boldsymbol{\gamma}\in\mathbb{R}^k$, denote the scaled difference in treatment and control group means of any linear combination of covariates. It follows that \(\mathbb{E}_{\text{lim}}\left[\sqrt{n}\widehat{\boldsymbol{\tau}}_{\boldsymbol{x}}\boldsymbol{\gamma}\mid D\leq a\right]=0\) and
		\begin{equation}
			\label{epvr}
			\frac{\operatorname{var}_{\text{lim}}\left(\sqrt{n}\widehat{\boldsymbol{\tau}}_{\boldsymbol{x}}\boldsymbol{\gamma}\right)-\operatorname{var}_{\text{lim}}\left(\sqrt{n}\widehat{\boldsymbol{\tau}}_{\boldsymbol{x}}\boldsymbol{\gamma}\mid D\leq a\right)}{\operatorname{var}_{\text{lim}}\left(\sqrt{n}\widehat{\boldsymbol{\tau}}_{\boldsymbol{x}}\boldsymbol{\gamma}\right)}=1-\nu_{a},
		\end{equation}
		where $\nu_{a}=\Pr(\chi_{k+2}^2\leq a)/\Pr(\chi_{k}^2\leq a)$ is a ratio of probabilities between two chi-square distributions of $k+2$ and $k$ degrees of freedom.
	\end{example}

	\section{Conditionally Affinely Invariant Rerandomization}
	\label{sec:main}
	\subsection{Motivation and Examples}
	Although affinely invariant rerandomization equally balances all covariates, in practice some covariates can be more predictive than others. \cite{morgan2015rerandomization} consider a situation where covariates vary in a priori importance, and suggest that better balance should be required for more important covariates. Another common situation is that covariates are a mixture of categorical and continuous variables. It is shown by \cite{johansson2022rerandomization} that stratification on binary covariates followed by rerandomization on continuous covariates is more efficient than rerandomization on all covariates at the same time.
	
	We first review balance criteria that address these concerns.
	\begin{example}[ReMT]
		\label{remt}
		\cite{morgan2015rerandomization} propose rerandomization using tiers of squared Mahalanobis distances (ReMT), where covariates are partitioned into \(T\) tiers (\(T > 1\)) ordered by decreasing importance. Covariates in each tier $t\geq2$ are projected onto those of lower tiers through
		\[
		\boldsymbol{x}^{(t)}=\boldsymbol{\beta}_0+\sum_{l=1}^{t-1}\boldsymbol{x}^{(l)}\boldsymbol{\beta}_{j}+\boldsymbol{e}^{(t)}
		\]
		and replaced with the projection residual $\boldsymbol{e}^{(t)}$, with $\boldsymbol{e}^{(1)}=\boldsymbol{x}^{(1)}$. For $1\leq t\leq T$, let $\widehat{\boldsymbol{\tau}}_{\boldsymbol{e}}^{(t)}=\overline{\boldsymbol{e}}^{(t)}_1 - \overline{\boldsymbol{e}}^{(t)}_0$ denote the difference in means for covariates in tier $t$ after the projection, and \(\operatorname{cov}\left(\widehat{\boldsymbol{\tau}}_{\boldsymbol{e}}^{(t)}\right)\) denote the sampling covariance of \(\widehat{\boldsymbol{\tau}}_{\boldsymbol{e}}^{(t)}\) under complete randomization. A randomization is rejected if the squared Mahalanobis distance for any tier \(t\),
		\(
		\widehat{\boldsymbol{\tau}}_{\boldsymbol{e}}^{(t)} \operatorname{cov}\left(\widehat{\boldsymbol{\tau}}_{\boldsymbol{e}}^{(t)}\right)^{-1} \widehat{\boldsymbol{\tau}}_{\boldsymbol{e}}^{(t)\top},
		\)
		exceeds a pre-specified threshold \(a_t > 0\).
	\end{example}
	In a stratified randomized experiment, units are divided in to $J$ strata according to the categorical variable $\boldsymbol{x}_i^{(s)}\in\{\boldsymbol{s}(1),\ldots,\boldsymbol{s}(J)\}$. Let $n_{j}$ denote the number of units in stratum $j$, with $n = \sum_{j=1}^J n_{j}$ the total units. A stratified randomization assigns units in each stratum to one of the two treatment arms through complete randomization \cite[see e.g.][Ch. 9]{imbens2015causal}. Although the categorical variable is balanced through stratification, there usually exist remaining covariates that need to be balanced.
	\begin{example}[SReM]
		\label{srem}
		To improve covariate balance in stratified experiments, \cite{wang2021rerandomization} propose two criteria for stratified rerandomization using squared Mahalanobis distances (SReM). For \(j = 1, \ldots, J\), let \(\widehat{\boldsymbol{\tau}}_{\boldsymbol{x}}^{(jr)}=\overline{\boldsymbol{x}}_{1}^{(jr)} - \overline{\boldsymbol{x}}_{0}^{(jr)}\) denote the difference in means of remaining covariates between treatment and control groups in stratum \(j\), and \(\operatorname{cov}\left(\widehat{\boldsymbol{\tau}}_{\boldsymbol{x}}^{(jr)}\right)\) denote its sampling covariance under stratified randomization. Across all strata, denote by \(\widehat{\boldsymbol{\tau}}_{\boldsymbol{x}} = n^{-1} \sum_{j=1}^J n_j  \widehat{\boldsymbol{\tau}}_{\boldsymbol{x}}^{(jr)}\), where \(n_j\) is the sample size in the $j$-th stratum. Let \(\operatorname{cov}\left(\widehat{\boldsymbol{\tau}}_{\boldsymbol{x}}\right)\) denote the sampling covariance of $\widehat{\boldsymbol{\tau}}_{\boldsymbol{x}}$ under stratified randomization. The first criterion is for overall balance, which rejects a randomization if the overall squared Mahalanobis distance,
		\(
		\widehat{\boldsymbol{\tau}}_{\boldsymbol{x}}\operatorname{cov}\left(\widehat{\boldsymbol{\tau}}_{\boldsymbol{x}}\right)^{-1} \widehat{\boldsymbol{\tau}}_{\boldsymbol{x}}^{\top},
		\)  
		exceeds a pre-specified threshold \(a>0\). The second criterion is for stratum-specific balance, which rejects a randomization if the squared Mahalanobis distance for any stratum \(j\),
		\(
		\widehat{\boldsymbol{\tau}}_{\boldsymbol{x}}^{(jr)} \operatorname{cov}\left(\widehat{\boldsymbol{\tau}}_{\boldsymbol{x}}^{(jr)}\right)^{-1}\widehat{\boldsymbol{\tau}}_{\boldsymbol{x}}^{(jr)\top},
		\)  
		exceeds a pre-specified threshold \(a_j > 0\).
	\end{example}
	
	When randomized experiments involve $Q>2$ treatment arms, covariate balance becomes an increasingly important issue \citep{branson2016improving}. We extend $\boldsymbol{W}$ as an $n\times (Q-1)$ matrix of mutually orthogonal dummy variables, where \(W_{iq} = 1\) if unit \(i\) is assigned to the \(q\)-th treatment arm (\(q = 1, \dots, Q-1\)), and \(0\) otherwise (units in the \(0\)-th arm have all entries \(0\)). Let \(n_q\) denote the sample size for arm \(q\), with \(\sum_{q=0}^{Q-1} n_q = n\). Specially, for \(2^K\) factorial experiments with \(K\) factors, each at levels \(+1\) and \(-1\), there are \(Q = 2^K\) treatment combinations. Factorial effects are defined by \(F = 2^K - 1\) orthogonal contrasts of potential outcomes \citep{dasgupta2015causal}. A completely randomized factorial design draws an assignment matrix randomly from $\mathcal{W}=\{\boldsymbol{W}\in\{0,1\}^{n\times(Q-1)}\mid\sum_{q=1}^{Q-1}W_{iq}\leq1\text{ for }1\leq i\leq n,\sum_{i=1}^{n}W_{iq}=n_q\text{ for }1\leq q\leq Q-1\}$.
	\begin{example}[ReFM]
		\label{frem}
		\cite{li2020rerandomization} consider rerandomization in factorial experiments with squared Mahalanobis distances (ReFM). Let $\overline{\boldsymbol{x}}=\left(\overline{\boldsymbol{x}}_0^{\top},\ldots,\overline{\boldsymbol{x}}_{Q-1}^{\top}\right)^{\top}$ denote the matrix of covariate means within $Q$ treatment arms. For the \(f\)-th factorial effect (\(1 \leq f \leq F\)), the contrast vector \(\boldsymbol{g}_f \in \{-1, +1\}^Q\) satisfying \(\sum_{q=1}^Q g_{fq} = 0\) defines
		\[
		\widehat{\boldsymbol{\tau}}_{\boldsymbol{x},f} = 2^{-(K-1)} \boldsymbol{g}_f \overline{\boldsymbol{x}},
		\]  
		which compares covariate means along this contrast. Let \(\widehat{\boldsymbol{\tau}}_{\boldsymbol{x}} = \left(\widehat{\boldsymbol{\tau}}_{\boldsymbol{x},1}, \dots, \widehat{\boldsymbol{\tau}}_{\boldsymbol{x},F}\right)\) denote the vector combining all factorial contrasts of covariates. Also, let \(\operatorname{cov}\left(\widehat{\boldsymbol{\tau}}_{\boldsymbol{x}}\right)\) denote the sampling covariance of $\widehat{\boldsymbol{\tau}}_{\boldsymbol{x}}$ under the completely randomized factorial design. \citet{li2020rerandomization} propose two rerandomization criteria for factorial designs. The first criterion rejects a randomization if the squared Mahalanobis distance, $\widehat{\boldsymbol{\tau}}_{\boldsymbol{x}}\operatorname{cov}\left(\widehat{\boldsymbol{\tau}}_{\boldsymbol{x}}\right)^{-1} \widehat{\boldsymbol{\tau}}_{\boldsymbol{x}}^{\top},$ exceeds a pre-specified threshold $a>0$. For the second criterion, factorial contrasts of covariates are partitioned into $H$ tiers. Let \(\widehat{\boldsymbol{\tau}}_{\boldsymbol{x}}^{[h]}\) denote the sub-vector of \(\widehat{\boldsymbol{\tau}}_{\boldsymbol{x}}\) containing factorial contrasts of covariates of tier \(h\) ($1\leq h\leq H$), and \(\widehat{\boldsymbol{\tau}}_{\boldsymbol{x}}^{[1:h-1]}\) denote the sub-vector of the previous $h-1$ tiers ($1< h\leq H$). Let $\operatorname{cov}\left(\widehat{\boldsymbol{\tau}}_{\boldsymbol{x}}^{[1:h-1]}\right)$ denote the sampling covariance of \(\widehat{\boldsymbol{\tau}}_{\boldsymbol{x}}^{[1:h-1]}\), and $\operatorname{cov}\left(\widehat{\boldsymbol{\tau}}_{\boldsymbol{x}}^{[h]},\widehat{\boldsymbol{\tau}}_{\boldsymbol{x}}^{[1:h-1]}\right)$ denote that between \(\widehat{\boldsymbol{\tau}}_{\boldsymbol{x}}^{[h]}\) and \(\widehat{\boldsymbol{\tau}}_{\boldsymbol{x}}^{[1:h-1]}\). For $h>1$, applying block-wise Gram-Schmidt orthogonalization, we obtain
		\begin{equation*}
			\widehat{\boldsymbol{\theta}}_{\boldsymbol{x}}^{[h]\top}= \widehat{\boldsymbol{\tau}}_{\boldsymbol{x}}^{[h]\top} - \operatorname{cov}\left(\widehat{\boldsymbol{\tau}}_{\boldsymbol{x}}^{[h]},\widehat{\boldsymbol{\tau}}_{\boldsymbol{x}}^{[1:h-1]}\right)\operatorname{cov}\left(\widehat{\boldsymbol{\tau}}_{\boldsymbol{x}}^{[1:h-1]}\right)^{-1}\widehat{\boldsymbol{\tau}}_{\boldsymbol{x}}^{[1:h-1]\top}
		\end{equation*}
		with \(\widehat{\boldsymbol{\theta}}_{\boldsymbol{x}}^{[ 1]}=\widehat{\boldsymbol{\tau}}_{\boldsymbol{x}}^{[ 1]}\). Let $\operatorname{cov}\left(\widehat{\boldsymbol{\theta}}_{\boldsymbol{x}}^{[h]}\right)$ denote the sampling covariance of $\widehat{\boldsymbol{\theta}}_{\boldsymbol{x}}^{[h]}$. The second criterion rejects a randomization if the squared Mahalanobis distance for any tier $h$, $\widehat{\boldsymbol{\theta}}_{\boldsymbol{x}}^{[h]}\operatorname{cov}\left(\widehat{\boldsymbol{\theta}}_{\boldsymbol{x}}^{[h]}\right)^{-1}\widehat{\boldsymbol{\theta}}_{\boldsymbol{x}}^{[h]\top},$ exceeds a pre-specified threshold $a_h>0$.
	\end{example}
	
	Balance criteria reviewed in Examples 2--4 can be effectively combined \cite[see e.g. rerandomization based on squared Mahalanobis distances of tiers of both covariates and factorial effects, ][]{li2018supplementary}, as they are built under a common principle introduced as follows.
	
	\subsection{Conditionally Affinely Invariant Criteria}
	\label{ssec:criteria}
	We now introduce the formal definition of conditionally affinely invariant rerandomization, which can be viewed as a close analogue to the conditionally affinely invariant matching \citep{rubin1992affinely}. Suppose that the covariate matrix is partitioned into tiers:
	\begin{equation}
		\label{partition}
		\boldsymbol{x}=\left(\boldsymbol{x}^{(s)}, \boldsymbol{x}^{(r)}\right)
	\end{equation}
	where $s,r$ stands for the special/remaining tiers of covariates, respectively.
	Let $\overline{\boldsymbol{x}}^{(r)} = \left( \overline{\boldsymbol{x}}_0^{(r)\top}, \ldots, \overline{\boldsymbol{x}}_{Q-1}^{(r)\top} \right)^{\top}$ denote the covariate mean vector for $Q \geq 2$ treatment arms within the remaining tier of $\boldsymbol{x}$. We analyze balance criteria depending on the scaled factorial contrasts:
	\begin{equation}
		\label{scaled_contrast}
		\sqrt{n}\, 2^{-(K-1)} \boldsymbol{g}_f \overline{\boldsymbol{x}}^{(r)},\quad 1 \leq f \leq F.
	\end{equation}
	When $Q=2$, this reduces to the scaled difference in means of remaining covariates $\sqrt{n}\left(\overline{\boldsymbol{x}}^{(r)}_1-\overline{\boldsymbol{x}}^{(r)}_0\right)$. When $Q > 2$, the set of scaled factorial contrasts can be further partitioned into special and remaining tiers as $ \{1,\ldots,F\}= \mathcal{F}_s\cup\mathcal{F}_r$, where $\mathcal{F}_s\cap\mathcal{F}_r=\emptyset$. Let $\boldsymbol{z}^{(r)}$ denote the scaled factorial contrasts in $\mathcal{F}_r$, and $\boldsymbol{z}^{(s)}$ denote those in $\mathcal{F}_s$ and other covariate tiers.
	
	We define the balance criterion $\varphi$ conditionally affinely invariant on $\boldsymbol{z}^{(r)}$ if it satisfies $\varphi\left(\mathcal{A}(\boldsymbol{z}^{(r)}),\boldsymbol{z}^{(s)}\right)=\varphi\left(\boldsymbol{z}^{(r)},\boldsymbol{z}^{(s)}\right)$. In Example \ref{remt}, by partitioning the column space of covariates into tiers, the balance criterion is conditionally affinely invariant on each tier. In Example \ref{srem}, by partitioning columns into strata and remaining covariates, both balance criteria are conditionally affinely invariant on remaining covariates. In Example \ref{frem}, the first criterion is affinely invariant, while the second one is conditionally affinely invariant on tiers of factorial contrasts.
	
	\subsection{The Conditional EPVR Property}
	We adopt the geometric definition of ellipsoidal distributions from \cite{rubin1992affinely}. A random vector \(\boldsymbol{X}\) is ellipsoidally distributed, denoted as \(\boldsymbol{X} \sim E(\boldsymbol{\mu}, \boldsymbol{\Sigma})\), if there exist a location parameter \(\boldsymbol{\mu}\) and a positive definite matrix \(\boldsymbol{\Sigma}\) such that
	\begin{equation}
		(\boldsymbol{X} - \boldsymbol{\mu})\boldsymbol{L}\boldsymbol{O} \sim (\boldsymbol{X} - \boldsymbol{\mu})\boldsymbol{L}
	\end{equation}
	for any orthogonal matrix \(\boldsymbol{O}\), where \(\boldsymbol{L}\) satisfies the Cholesky decomposition \(\boldsymbol{\Sigma}^{-1} = \boldsymbol{L}\boldsymbol{L}^{\top}\).
	
	The family extends to conditionally ellipsoidal distributions by relaxing rotation symmetry in some dimensions. Consider a partition of \(\boldsymbol{X}\) as
	\(
	\boldsymbol{X} = \left(\boldsymbol{X}^{(r)}, \boldsymbol{X}^{(s)}\right)
	\).
	We say \(\boldsymbol{X}\) is conditionally ellipsoidally distributed if
	\begin{equation*}
		\boldsymbol{X}^{(r)} \mid \boldsymbol{X}^{(s)} \sim E\left(\boldsymbol{\mu}_{\boldsymbol{X}^{(s)}}, \boldsymbol{\Sigma}_r\right),
	\end{equation*}
	where \(\boldsymbol{\mu}_{\boldsymbol{X}^{(s)}}\) is a location parameter dependent on \(\boldsymbol{X}^{(s)}\), and \(\boldsymbol{\Sigma}_r\) is a fixed positive definite matrix. The canonical form \(\left(\boldsymbol{C}^{(r)}, \boldsymbol{X}^{(s)}\right)\) is obtained by \begin{equation}
		\label{canonical}
		\boldsymbol{C}^{(r)}=\left(\boldsymbol{X}^{(r)}-\boldsymbol{\mu}_{\boldsymbol{X}^{(s)}}\right)\boldsymbol{L}_r, 
	\end{equation}
	where \(\boldsymbol{L}_r\) satisfies the Cholesky decomposition \(\boldsymbol{\Sigma}_r^{-1} = \boldsymbol{L}_r\boldsymbol{L}_r^{\top}\). By construction, $\boldsymbol{C}^{(r)} \mid \boldsymbol{X}^{(s)} \sim E(\boldsymbol{0}, \boldsymbol{I})$,
	where \(\boldsymbol{0}\) is a zero vector and \(\boldsymbol{I}\) is an identity matrix.

	Suppose that $\left(\boldsymbol{z}^{(r)}, \boldsymbol{z}^{(s)}\right)$ has a limiting distribution that is conditionally ellipsoidally symmetric. Through the affine transformation \eqref{canonical}, $\left(\boldsymbol{c}^{(r)}, \boldsymbol{z}^{(s)}\right)$ converges in distribution to its canonical form. For balance criterion $\varphi$ conditionally affinely invariant on $\boldsymbol{z}^{(r)}$, it implies that $\varphi\left(\boldsymbol{z}^{(r)}, \boldsymbol{z}^{(s)}\right) = \varphi\left(\boldsymbol{c}^{(r)}, \boldsymbol{z}^{(s)}\right)$ with $\varphi$ retaining conditional affine invariance on $\boldsymbol{c}^{(r)}$.
	The next theorem establishes the conditionally EPVR property of such balance criteria.
	
	\begin{theorem}
		\label{shrink_cond}
		As $n \to \infty$, suppose that $\left(\boldsymbol{c}^{(r)}, \boldsymbol{z}^{(s)}\right)$ has a limiting distribution that is conditionally ellipsoidally symmetric in the canonical form. If $\varphi$ is conditionally affinely invariant on $\boldsymbol{c}^{(r)} $ and almost surely continuous, then
		\begin{align}
			\label{t1}
			\mathbb{E}_{\text{lim}}\left[\boldsymbol{c}^{(r)}  \mid \varphi = 1\right] &= \boldsymbol{0}, \\
			\label{t2}
			\operatorname{cov}_{\text{lim}}\left(\boldsymbol{c}^{(r)}  \mid \varphi = 1\right) &\propto\boldsymbol{I}.
		\end{align}
	\end{theorem}
	
	\begin{proof}
		We use $\dot{\sim}$ to denote convergence to the same distribution. By the continuous mapping theorem, the asymptotic distribution satisfies
		\[
		\left(\boldsymbol{c}^{(r)} ,   \boldsymbol{z}^{(s)}\right) \dot{\sim} \left(\boldsymbol{c}^{(r)} \boldsymbol{O},   \boldsymbol{z}^{(s)}\right),
		\]
		where $\boldsymbol{O}$ is any orthogonal matrix. The invariance of $\varphi$ under affine transformations on $\boldsymbol{c}^{(r)}$ and the almost sure continuity of $\varphi$ imply
		\[
		\left(\boldsymbol{c}^{(r)} ,   \boldsymbol{z}^{(s)}, \varphi\right) \dot{\sim} \left(\boldsymbol{c}^{(r)} \boldsymbol{O}_r,   \boldsymbol{z}^{(s)}, \varphi\right).
		\]
		Therefore, $\left(\boldsymbol{c}^{(r)}, \left(\boldsymbol{z}^{(s)},\varphi\right)\right)$ has a limiting distribution that is conditionally ellipsoidally symmetric in the canonical form, which leads to the following calculation of moments:
		\begin{align*}
			\mathbb{E}_{\text{lim}}\left[\boldsymbol{z}^{(r)}\mid\boldsymbol{z}^{(s)}, \varphi = 1\right] &= \boldsymbol{0}, \\
			\operatorname{cov}_{\text{lim}}\left(\boldsymbol{z}^{(r)}\mid\boldsymbol{z}^{(s)}, \varphi = 1\right) &\propto\boldsymbol{I}.
		\end{align*}
		Taking expectations over $\boldsymbol{z}^{(s)}$ completes the proof.
	\end{proof}
	For any $\boldsymbol{\gamma}  \in \mathbb{R}^{k_r|\mathcal{F}_r|}$, Theorem \ref{shrink_cond} implies that
	\begin{equation}
		\label{epvr_2k}
		\frac{\operatorname{var}_{\text{lim}}\left(\boldsymbol{c}^{(r)} \boldsymbol{\gamma} \right) - \operatorname{var}_{\text{lim}}\left(\boldsymbol{c}^{(r)} \boldsymbol{\gamma}  \mid \varphi = 1\right)}{\operatorname{var}_{\text{lim}}\left(\boldsymbol{c}^{(r)} \boldsymbol{\gamma} \right)} = \text{constant}.
	\end{equation}
	This conditionally EPVR property generalizes \eqref{epvr}, resulting from the local symmetry of balance criterion and covariate limiting distribution.
	
	A statistical decision problem can be raised as whether the asymptotic variance reduction can be further improved without tightening the asymptotic acceptance probability. Existing frameworks either lack a positive lower bound on the acceptance probability for valid randomization-based inference \cite[e.g.,][]{kallus2018optimal}, or depends on prior information \cite[e.g.,][]{lu2023design,liu2023bayesian} which is often unavailable in practice. Within our framework introduced as follows, a new class of balance criteria should be favored if it dominates existing ones on selected balance metrics.
	
	\section{A Decision-Theoretic Framework for Evaluating Balance Criteria}
	\label{sec:decision_theory}
	
	To address the statistical decision problem, we first extend the definition of balance criteria. Let \(\varphi: \mathcal{W} \to [0,1]\) be a stochastic balance criterion, where \(\varphi(\omega)\) defines the probability of accepting a randomization \(\omega\). The decision to accept the randomization when $\delta(\omega)=1$ and reject otherwise follows a Bernoulli distribution with \(\Pr(\delta(\omega)=1\mid\omega)=\varphi(\omega)\). The space \(\mathcal{C}\) contains all stochastic balance criteria, with \(\Pr(\delta=1)=\mathbb{E}[\varphi] \geq p>0\) ensuring valid randomization-based inference.
	
	Let \(\zeta: \mathcal{W} \to \mathbb{R}\) quantify covariate balance (e.g., the squared Mahalanobis distance). By the law of total expectation, conditional on accepted samples, we have
	\[
	\mathbb{E}[\zeta\mid\delta=1] = \frac{\mathbb{E}[\zeta \delta]}{\mathbb{E}[\delta]} = \frac{\mathbb{E}[\zeta \varphi]}{\mathbb{E}[\varphi]}:=\mathbb{E}[\zeta \mid \varphi].
	\]
	Given multiple imbalance measures \(\{\zeta_\theta : \theta \in \Theta\}\), a criterion \(\varphi \in \mathcal{C}\) is inadmissible if it can be dominated by another $\varphi'\in\mathcal{C}$ such that
	\begin{equation*}
		\mathbb{E}[{\zeta}_{\theta}\mid\varphi']\leq\mathbb{E}[{\zeta}_{\theta}\mid\varphi],\ \theta\in\Theta,
	\end{equation*}
	and that at least one inequality is strict.
	A criterion is admissible if no such \(\varphi'\) exists. A class \(\mathcal{C}^* \subseteq \mathcal{C}\) is complete if every \(\varphi \notin \mathcal{C}^*\) is dominated by some \(\varphi \in \mathcal{C}^*\), and is admissible complete if, additionally, every \(\varphi \in \mathcal{C}^*\) is admissible. These concepts parallel to the classical decision theory \citep[see e.g.][]{berger1985statistical}, but with non-convex objectives \(\mathbb{E}[\zeta_\theta \mid \varphi]\). A new complete class theorem is provided to characterize admissible/inadmissible balance criteria with respect to a finite number of conditional expectations.
	
	\begin{theorem}
		\label{complete_class}
		Suppose that $\Theta=\{1,2,\ldots,M\}$. Let $\mathcal{C}^{*}$ contain all $\varphi\in\mathcal{C}$ that satisfies the necessary condition below.
		
		\noindent$(\romannumeral1)$ A sufficient condition for $\varphi\in\mathcal{C}$ with $\mathbb{E}[\varphi]=p$ to be admissible is the existence of $a\in\mathbb{R}$ and positive weights $\pi_1,\ldots,\pi_M$ such that
		\begin{equation}
			\label{adm_cond}
			\varphi(\omega)=
			\begin{cases}
				1,& \sum_{m=1}^{M}\pi_{m}\zeta_{m}(\omega)<a, \\
				0,& \sum_{m=1}^{M}\pi_{m}\zeta_{m}(\omega)>a.
			\end{cases}
		\end{equation}
		
		\noindent$(\romannumeral2)$ A necessary condition for $\varphi\in\mathcal{C}$ to be admissible is the existence of $a\in\mathbb{R}$ and nonnegative weights $\pi_1,\ldots,\pi_m$ such that (\ref{adm_cond}) holds almost surely.
		
		\noindent$(\romannumeral3)$ $\mathcal{C}^*$  is complete.
	\end{theorem}
	
	We focus on two sub-classes of balance criteria frequently applied in practice. A unified criterion places a threshold on a single weighted imbalance measure:
	\[
	I\left(\sum_{m=1}^M \pi_m \zeta_m(\omega) \leq a\right),\quad\pi_m \geq 0,\ a \in \mathbb{R},\ 1\leq m\leq M.
	\]
	In contrast, an intersection criterion sequentially places thresholds on the $m$-th objective:
	\[
	I\left(\zeta_m(\omega) \leq a_m\right),\quad a_m\in\mathbb{R},\ 1\leq m\leq M.
	\]
	Conceptually, the intersection criterion is equivalent to
	\(
	\prod_{m=1}^{M}I\left(\zeta_m(\omega)\leq a_m\right).
	\)
	However, as the process can be terminated before the last step, the cost of making a decision can be reduced.
	
	\begin{proposition}
		\label{prop:cost}
		For $1\leq m\leq M$, let \(h_m(\omega)\) be the cost of collecting and storing \(\zeta_m(\omega)\), which is assumed to be non-negative and integrable. For an intersection criterion with \(M\) steps, the expected cost is no larger than $\sum_{m=1}^{M}\mathbb{E}[h_m]$, the average cost of making a decision until the final step.
	\end{proposition}
	For covariates available at the design stages in RCTs, the cost of rerandomization is computing and storing balance scores, which is not a major concern. While intersection methods may reduce computational cost by early termination, they are generally inadmissible. We illustrate this point with a class of popular balance criteria, the covariate balance table check.
	\begin{example}[ReP]
		\label{balance_table}
		Contemporary RCTs often include covariate balance tables to check randomization quality. According to a survey by \cite{bruhn2009pursuit}, practitioners tend to rerandomize if the standard $t$-test shows significant results on some covariates. \cite{zhao2024no} formalize this practice as rerandomization using marginal rules (ReP). For each covariate \(m = 1, \ldots, k\), define the standardized difference  
		\[
		t_{m} = \frac{\overline{\boldsymbol{x}}^{(m)}_1 - \overline{\boldsymbol{x}}^{(m)}_0}{\sqrt{\frac{(n_1 - 1)s_{1}^{(m)} + (n_0 - 1)s_{0}^{(m)}}{n - 2} \left( \frac{1}{n_1} + \frac{1}{n_0} \right)}},
		\]
		where \(\overline{\boldsymbol{x}}^{(m)}_z\) and \(s_z^{(m)}\) are the sample mean and variance of the \(m\)-th covariate in treatment arm \(z \in \{0, 1\}\). Let \(a_1, \ldots, a_k \geq 0\) denote pre-specified thresholds. The balance criterion \(\prod_{m=1}^k I(|t_m| \leq a_m)=I\left(|\boldsymbol{t}| \leq \boldsymbol{a}\right)\) rejects randomizations where any \(|t_m|\) exceeds \(a_m\).
		
		Let \(\boldsymbol{S}_{\boldsymbol{x}}=\frac{1}{n-1}\sum_{i=1}^{n}\left(\boldsymbol{x}_i-\overline{\boldsymbol{x}}\right)^{\top}\left(\boldsymbol{x}_i-\overline{\boldsymbol{x}}\right)\) denote the sample covariance matrix of covariates, and \(\operatorname{diag}(\boldsymbol{S}_{\boldsymbol{x}})\) the diagonal matrix of \(\boldsymbol{S}_{\boldsymbol{x}}\). \cite{zhao2024no} show that, under certain regularity conditions, \(\boldsymbol{t} = (t_1, \ldots, t_k)\) converges in distribution to \(N(\boldsymbol{0}, \operatorname{diag}(\boldsymbol{S}_{\boldsymbol{x}})^{-1} \boldsymbol{S}_{\boldsymbol{x}} \operatorname{diag}(\boldsymbol{S}_{\boldsymbol{x}})^{-1})\), and that the asymptotic variances of Neyman's or Fisher's estimators under ReP decomposes into \(\operatorname{var}_{\text{lim}}\left(\boldsymbol{t\gamma}\mid |\boldsymbol{t}| \leq \boldsymbol{a}\right)\) for some $\boldsymbol{\gamma}\in\mathbb{R}^{k}$ and an error term no affected by ReP.
		A statistical question can be raised as whether the asymptotic variance \(\operatorname{var}_{\text{lim}}\left(\boldsymbol{t\gamma}\mid |\boldsymbol{t}| \leq \boldsymbol{a}\right)\) can be further reduced for any fixed $\boldsymbol{\gamma}\in\mathbb{R}^{k}$ without tightening \(\lim_{n\rightarrow\infty}\Pr\left(|\boldsymbol{t}| \leq \boldsymbol{a}\right)\).
		
		Note that the asymptotic variance under $I(|\boldsymbol{t}| \leq \boldsymbol{a})$ can be decomposed as
		\begin{equation}
			\label{variance_decomp}
			\begin{aligned}
				&\operatorname{var}_{\text{lim}}\left(\boldsymbol{t\gamma}\mid |\boldsymbol{t}| \leq \boldsymbol{a}\right)=\sum_{m=1}^{k}\gamma_{m}^2\mathbb{E}_{\text{lim}}\left[t_m^2\mid|\boldsymbol{t}| \leq \boldsymbol{a}\right]\\+&2\sum_{m=1}^{k}\sum_{\ell>m}\left(\gamma_{\ell}\gamma_{m}\right)^+\mathbb{E}_{\text{lim}}\left[t_{\ell}t_m\mid|\boldsymbol{t}| \leq \boldsymbol{a}\right]+2\sum_{m=1}^{k}\sum_{\ell<m}\left(\gamma_{\ell}\gamma_{m}\right)^-\mathbb{E}_{\text{lim}}\left[-t_{\ell}t_m\mid|\boldsymbol{t}| \leq \boldsymbol{a}\right],
			\end{aligned}
		\end{equation}
		where $\left(\gamma_{\ell}\gamma_{m}\right)^+$ denotes the maximum of $\gamma_{\ell}\gamma_{m}$ and $0$, and $\left(\gamma_{\ell}\gamma_{m}\right)^-$ the minimum of $\gamma_{\ell}\gamma_{m}$ and $0$. We consider \(k^2\) integrable functions:
		\begin{equation}
			\label{targets}
			\begin{aligned}
				&t_m^2 ,\ 1\leq m\leq k,\\
				&t_\ell t_m, \ 1\leq m<\ell\leq k,\\
				-&t_\ell t_m, \ 1\leq\ell<m\leq k.
			\end{aligned}
		\end{equation}
		Let $\zeta=\sum_{m=1}^k \left(\pi_{mm} t_m^2 + \sum_{m<\ell\leq k} \pi_{\ell m} t_\ell t_m - \sum_{1\leq\ell<m} \pi_{\ell m} t_\ell t_m\right)$ be any non-negative weighted sum of these functions.
		By Theorem \ref{complete_class}, in terms of minimizing the asymptotic conditional expectations of \eqref{targets}, the following class of balance criteria
		\[
		\mathcal{C}^*=\left\{\varphi=I(\zeta\leq a),\text{a.s.}\mid\lim_{n\rightarrow\infty}\Pr(\zeta\leq a)\geq\lim_{n\rightarrow\infty}\Pr(|\boldsymbol{t}| \leq \boldsymbol{a})\right\}
		\]
		is complete. However, \(I(|\boldsymbol{t}| \leq \boldsymbol{a})\) lies outside $\mathcal{C}^*$, as the acceptance region of each balance criterion of $\mathcal{C}^*$ is defined by a quadratic form up to almost sure equivalence. Also, the asymptotic variance under $I(\zeta\leq a)$ admits the decomposition \eqref{variance_decomp}. We conclude that \(I(|\boldsymbol{t}| \leq \boldsymbol{a})\) is inadmissible and can be dominated by some $I(\zeta\leq a)\in\mathcal{C}^*$ such that \[\operatorname{var}_{\text{lim}}\left(\boldsymbol{t\gamma}\mid \zeta\leq a\right)\leq\operatorname{var}_{\text{lim}}\left(\boldsymbol{t\gamma}\mid |\boldsymbol{t}| \leq \boldsymbol{a}\right)\] for any $\boldsymbol{\gamma}\in\mathbb{R}^k$ with at least one inequality being strict.
	\end{example}
	
	\section{Rerandomization Based on Weighting Mahalanobis Distances}
	\label{sec:rewm}
	In Example \ref{balance_table}, balance criteria within the complete class may not be conditionally EPVR. Moreover, specifying $k^2$ weights becomes impractical with high-dimensional covariates. Theorem \ref{shrink_cond} shows that, by imposing conditional affine invariance on balance criteria, the conditionally EPVR property holds within invariant subspaces, which reduces weight selection for imbalance measures structured as Mahalanobis distances.
	
	We develop a unified framework to construct balance criteria for experimental designs involving tiers of covariates, stratification, and multiple treatment arms. To evaluate the improvement in causal estimation, we adopt a potential outcome framework from \cite{dasgupta2015causal}. For $Q\geq2$, let $Y_i(q)$ denote the potential outcome for unit $i$ under the $q$-th treatment ($0\leq q\leq Q-1$), and let $\boldsymbol{Y}_i=\left(Y_i(0),\ldots,Y_i(Q-1)\right)$ denote the vector of all potential outcomes. For $1\leq f\leq F$, the $f$-th factorial effect for each unit $i$ is defined as $\tau_{fi}=2^{-(K-1)}\boldsymbol{g}_f\boldsymbol{Y}_i$, with the average factorial effect as $\tau_{f}=n^{-1}\sum_{i=1}^{n}\tau_{fi}$.
	Let $y_i$ denote the observed outcome of unit $i$ and $\boldsymbol{y}$ the observed outcome vector for all units under the realized treatment assignment $\boldsymbol{W}$. For each factorial effect, we consider $\widehat{\tau}_f=2^{-(K-1)}\boldsymbol{g}_f\overline{\boldsymbol{y}}$ to estimate $\tau_{f}$, where $\overline{\boldsymbol{y}}=\left(\overline{y}_0,\ldots,\overline{y}_{Q-1}\right)^{\top}$ denotes the vector of observed outcome means of $Q\geq2$ treatment arms. When $Q=2$, it reduces to the standard difference-in-means estimator $\overline{y}_{1}-\overline{y}_{0}$ for the average treatment effect. Denote by $\widehat{\boldsymbol{\tau}}=(\widehat{\tau}_1,\ldots,\widehat{\tau}_F)$ and $\boldsymbol{\tau}=(\tau_1,\ldots,\tau_F)$. Recall the definition of $\widehat{\boldsymbol{\tau}}_{\boldsymbol{x}}$ in Example \ref{frem}. \cite{li2017general} establish general forms of finite population central limit theorems (CLTs), showing that under certain regularity conditions,
	\begin{equation}
		\label{joint_normality}
		\left(\sqrt{n}(\widehat{\boldsymbol{\tau}}-\boldsymbol{\tau}),\sqrt{n}\widehat{\boldsymbol{\tau}}_{\boldsymbol{x}}\right)\stackrel{d}{\rightarrow}N(\boldsymbol{0},\boldsymbol{V}),\text{ as } n\rightarrow\infty,
	\end{equation}
	where $\boldsymbol{V}$ is the limit of the joint sampling covariance of $\left(\sqrt{n}(\widehat{\boldsymbol{\tau}}-\boldsymbol{\tau}),\sqrt{n}\widehat{\boldsymbol{\tau}}_{\boldsymbol{x}}\right)$. \cite{li2017general} also note that for stratified experiments with few strata but large within-stratum sample sizes, the finite population CLT can be applied independently to each stratum, and averaging across strata also leads to \eqref{joint_normality}. For the case with many small strata, \cite{liu2024randomization} establish \eqref{joint_normality} using an extended finite population CLT.
	
	We partition factorial contrasts into $H\geq1$ tiers and covariates into $T\geq1$ tiers based on their importance. Following Section \ref{ssec:criteria}, we construct $M=H\times T$ scaled linear contrast vectors $\{\boldsymbol{z}^{(m)}\}_{m=1}^M$. Then $(\boldsymbol{z}^{(1)},\ldots,\boldsymbol{z}^{(M)})$ is a permutation of $\sqrt{n}\widehat{\boldsymbol{\tau}}_{\boldsymbol{x}}$.
	To orthogonalize tiers of contrasts, for $1<m\leq M$, we apply the block-wise Gram-Schmidt orthogonalization to obtain the residual $\widetilde{\boldsymbol{z}}^{(m)}$ by projecting $\boldsymbol{z}^{(m)}$ onto $\left(\boldsymbol{z}^{(1)},\ldots,\boldsymbol{z}^{(m-1)}\right)$ with $\widetilde{\boldsymbol{z}}^{(1)}=\boldsymbol{z}^{(1)}$. The transformation from $(\boldsymbol{z}^{(1)},\ldots,\boldsymbol{z}^{(M)})$ to $\left(\widetilde{\boldsymbol{z}}^{(1)},\ldots,\widetilde{\boldsymbol{z}}^{(M)}\right)$ is based on the sampling covariance, which has a limit $\boldsymbol{V}$ as $n\rightarrow\infty$. Applying the continuous mapping theorem, the joint asymptotic normality \eqref{joint_normality} implies
	\begin{equation}
		\label{joint_normality_block}
		\left(\widetilde{\boldsymbol{z}}^{(1)},\ldots,\widetilde{\boldsymbol{z}}^{(M)}\right)\ \dot{\sim}\ N(\boldsymbol{0},\operatorname{diag}\left(\operatorname{cov}\left(\widetilde{\boldsymbol{z}}^{(1)}\right),\ldots,\operatorname{cov}\left(\widetilde{\boldsymbol{z}}^{(M)}\right)\right)),
	\end{equation}
	where $\operatorname{cov}\left(\widetilde{\boldsymbol{z}}^{(m)}\right)$ is the sampling covariance of $\widetilde{\boldsymbol{z}}^{(m)}$ $(1\leq m\leq M)$.
	
	We now introduce the squared Mahalanobis distance for each $m$-th orthogonalized contrast as 
	\[
	D_m =\widetilde{\boldsymbol{z}}^{(m)}\operatorname{cov}\left(\widetilde{\boldsymbol{z}}^{(m)}\right)^{-1}\widetilde{\boldsymbol{z}}^{(m)\top}.
	\]
	Through Cholesky decomposition $\operatorname{cov}\left(\widetilde{\boldsymbol{z}}^{(m)}\right)^{-1}=\boldsymbol{L}_m\boldsymbol{L}_m^{\top}$, we define the canonical contrast as $\boldsymbol{c}^{(m)}=\widetilde{\boldsymbol{z}}^{(m)}\boldsymbol{L}_m$, and write $D_m=\Vert\boldsymbol{c}^{(m)}\Vert^2$, where $\Vert\cdot\Vert$ is the Euclidean distance in $\mathbb{R}^{k_m}$.
	Note that any balance criterion conditionally affinely invariant on \(\boldsymbol{z}^{(m)}\) is equivalent to that invariant on \(\boldsymbol{c}^{(m)}\). 
	Let $\mathcal{C}$ denote the class of balance criteria with asymptotic acceptance probabilities bounded below by 
	\begin{equation}
		\lim_{n \to \infty} \Pr(\varphi = 1) \geq p > 0,\quad\varphi\in\mathcal{C}.
	\end{equation}
	Let \(\mathcal{C}_M\subset\mathcal{C}\) denote the subclass defined on $\left(\boldsymbol{c}_1,\ldots,\boldsymbol{c}_M\right)$ that is almost surely continuous and conditionally affinely invariant on each \(\boldsymbol{c}^{(m)}\).
	The subclass \(\mathcal{C}_M^* \subset \mathcal{C}_M\) defines rerandomization with weighted Mahalanobis distances (ReWM), which consists of unified balance criteria in the form:
	\begin{equation*}
		\varphi^*=I\left( \sum_{m=1}^M \pi_m D_m \leq a \right), \quad \pi_m \geq 0, \,\, a > 0.
	\end{equation*}
	From \eqref{joint_normality_block}, we have
	\begin{equation*}
		(D_1,\ldots,D_M)\ \dot{\sim}\ \left(\chi_{k_1}^2,\ldots,\chi_{k_M}^2\right),
	\end{equation*}
	where each $\chi_{k_m}^2$ is a chisquare random variable of $k_m$-degree of freedom independent of the rest. In practice, we can draw random samples from $\left(\chi_{k_1}^2,\ldots,\chi_{k_M}^2\right)$ to set $a$ as the $p$-th quantile of $\sum_{m=1}^{M}\pi_m\chi_{k_m}^2$.
	
	Consider a general scalar estimand $ \tau=\sum_{f=1}^{F}\lambda_f\tau_f$ defined as a convex combination of factorial effects, which can be estimated by $\widehat{\tau}=\sum_{f=1}^{F}\lambda_f\widehat{\tau}_f$ for pre-specified $\lambda_1,\ldots,\lambda_F\in [0,1]$. The following standardized estimator decomposes into projections on canonical contrasts and a residual term $c_0$ uncorrelated with these contrasts:
	\[
	v^{-1/2}\left(\widehat{\tau}- \tau\right) = \sum_{m=1}^M \boldsymbol{c}^{(m)} \boldsymbol{\gamma}_m + c_0,
	\]
	where $v=\operatorname{var}(\widehat{\tau})$ is the sampling variance of $\widehat{\tau}$ under complete randomization, and each \(\boldsymbol{\gamma}_m\) is the projection coefficient. The squared multiple correlation between $v^{-1/2}\left(\widehat{\tau}- \tau\right)$ and $\boldsymbol{c}^{(m)}$ is \(\rho_m^2 = \Vert\boldsymbol{\gamma}_m\Vert^2\), representing the proportion of variance explained by the \(m\)-th canonical contrast. Under \eqref{joint_normality}, the standardized estimator has the limiting distribution:
	\[
	v^{-1/2}\left(\widehat{\tau}- \tau\right)\ \dot{\sim}\ \sum_{m=1}^M \boldsymbol{\epsilon}^{(m)} \boldsymbol{\gamma}_m + \sqrt{1 - \sum_{m=1}^M \rho_m^2} \cdot \epsilon_0,
	\]
	where \(\boldsymbol{\epsilon}^{(m)} \sim N(\boldsymbol{0}, \boldsymbol{I})\) and \(\epsilon_0 \sim N(0, 1)\) are independent. The rotation symmetry in each $\boldsymbol{\epsilon}^{(m)}$ further implies that
	\[
	v^{-1/2}\left(\widehat{\tau}- \tau\right)\ \dot{\sim}\ \sum_{m=1}^M \rho_m\epsilon_1^{(m)} + \sqrt{1 - \sum_{m=1}^M \rho_m^2} \cdot \epsilon_0,
	\]
	where $\epsilon_1^{(m)}$ is the first component of $\boldsymbol{\epsilon}^{(m)}$.

	\begin{corollary}
		\label{cor:epvr}  
		Assume \eqref{joint_normality} holds and \(\varphi \in \mathcal{C}_M\).\\
		\noindent$(\romannumeral1)$ The standardized estimator satisfies
		\begin{equation}
			\label{pos_rewm}
			v^{-1/2}\left(\widehat{\tau} -  \tau\right) \mid \{\varphi=1\} \ \dot{\sim}\  \sum_{m=1}^M \rho_m\epsilon_1^{(m)} \mid \{\varphi=1\} + \sqrt{1 - \sum_{m=1}^M \rho_m^2} \cdot \epsilon_0
		\end{equation}
		with the asymptotic variance decomposes as
		\begin{equation}
			\label{var_decomp}
			\operatorname{var}_{\text{lim}}\left(v^{-1/2}\widehat{\tau} \mid \varphi = 1\right) = \sum_{m=1}^M \frac{\rho_m^2}{k_m}\mathbb{E}_{\text{lim}}[D_m \mid \varphi=1] + \left(1 - \sum_{m=1}^M \rho_m^2\right).
		\end{equation}
		\noindent$(\romannumeral2)$ \(\mathcal{C}_M^*\) is an admissible complete class for minimizing \(\{\mathbb{E}_{\text{lim}}[D_m \mid \varphi=1]\}_{m=1}^M\).\\
		\noindent$(\romannumeral3)$ Any intersection criterion $\prod_{m=1}^M I(D_m \leq a_m)\in\mathcal{C}_M$ is inadmissible in the same sense as (ii) for $M>1$.
	\end{corollary}
	
	By Corollary \ref{cor:epvr}, there is no balance criterion in \(\mathcal{C}_M\) can achieves a smaller asymptotic variance than $\varphi^*\in\mathcal{C}_M^*$ for any standardized estimator with fixed \(\boldsymbol{\rho}=(\rho_1, \ldots, \rho_M)\). Moreover, ReWM dominates intersection methods reviewed in Examples \ref{rem}--\ref{frem} for balancing contrasts. Specifically, ReMT is inadmissible when $T>1$, SReM using the second balance criterion is inadmissible when $J>1$, and ReFM using the second balance criterion is inadmissible when $H>1$.
	
	By considering $\sum_{m=1}^{M}\frac{\rho_m^2}{k_m}D_m$ as the balance metric, Theorem \ref{complete_class} also implies the asymptotic optimal balance criterion in $\mathcal{C}_M^{*}$ for minimizing $\operatorname{var}_{\text{lim}}\left(v^{-1/2}\widehat{\tau} \mid \varphi = 1\right)$ to be
	\begin{equation*}
		I\left(\sum_{m=1}^{M}\frac{\rho_m^2}{k_m}D_m\leq a\right),
	\end{equation*}
	provided that \(\boldsymbol{\rho}\) is known. Although \(\boldsymbol{\rho}\) is in fact unknown in practice, this result suggest a way of choosing weights. Given a collection of estimators $\widehat{\tau}_1,\ldots,\widehat{\tau}_F$ for factorial effects, we firstly assess their importance with weights $\widetilde{\pi}_{f}$, $1\leq f\leq F$. We then assess the squared multiple correlations $\widetilde{\rho}_{mf}^2$ that how much $\widetilde{\boldsymbol{z}}^{(m)}$ can explain $\widehat{\tau}_f$. The final weights aggregates across tiers of factorial effects as
	\begin{equation}
		\label{weight_choice}
		\pi_m\propto\sum_{f=1}^{F}\widetilde{\pi}_f\frac{\widetilde{\rho}_{mf}^2}{k_m}.
	\end{equation}
	
	In Supplementary Material, we provide more on Neymanian inference to construct the asymptotically conservative interval for the finite population estimand $\tau$. In another paper devoted to Fisherian inference \citep{zhong2025fisher}, we discuss the application of randomization tests to complex designs like rerandomization, which enable inductive inference about causal estimands for larger populations.
	
	\section{A Realistic Example}
	\label{sec:example}
	\cite{blattman2017reducing} conducted a stratified factorial experiment to evaluate the impact of noncognitive skills and preferences on reducing crime and violence. In their design, participants were assigned to four arms: cognitive behavioral therapy (CBT) alone, unconditional \$200 cash transfers alone, combined CBT-cash, and a control group, where the CBT program taught participants to reduce impulsive behaviors and adopt prosocial identities through eight weeks of group sessions.
	
	The balance tables in \cite{blattman2017reducing} reveal that, of $57$ covariates over $3$ treatment arms, $14$ ($8.2$ percent) are found to be significantly imbalanced with $p<0.05$. The authors use auxiliary analyses to justify that baseline imbalances are unlikely to be driving their results, and control for baseline covariates in their main empirical specification. With rerandomization, we can circumvent the covariate imbalance in the design stage, thereby improving the quality of the study.
	
	We divide factorial contrasts into 2 tiers: main effects for tier 1 and the interaction effect for tier 2. We divide covariates and their combinations into 3 tiers. The first tier contains $4$ prognostically important covariates including Age, Year of Schooling, Cognitive Ability and Summary Index of Income. The second tier contains the remaining $53$ covariates. The third tier contains the $14$ interaction and quadratic terms of the covariates in tier 1. We then construct 6 tiers of scaled factorial contrasts of covariates, and order them by tiers of effects and tiers of covariates.
	
	We equally weight 3 factorial contrasts, and weight 3 tiers of covariates as $1:1:0.6$ to reflect their ability to explain factorial effects. Using \eqref{weight_choice} to determine weights, the ReWM criterion can be given by
	\begin{equation*}
		\varphi_1=I\left(\frac{1}{8}D_1+\frac{1}{106}D_2+\frac{0.6}{28}+\frac{0.5}{4}D_4+\frac{0.5}{53}D_5+\frac{0.3}{14}D_6\leq a\right),
	\end{equation*}
	where $a$ is chosen to have $\lim_{n\rightarrow\infty}\Pr(\varphi_1=1)=0.001$.
	In comparison, we set up an intersection balance criterion:
	\begin{equation*}
		\varphi_2=\prod_{m=1}^{6}I(D_i\leq a_m),
	\end{equation*}
	where $a_1,\ldots,a_6$ are chosen to be the $0.1$-th, $0.25$-th, $0.5$-th, $0.2$-th, $0.5$-th and $0.8$-th quantiles of the chi-square distributions of $8,106,28,4,53$ and $14$ degrees of freedoms, respectively. This ensures that $\lim_{n\rightarrow\infty}\Pr(\varphi_2=1)=0.001$.
	
	The original experiment uses two separate block designs for treatment assignment. Focusing on the first blocking variable, we implement a stratified completely randomized factorial design. We rerandomize 1,000,000 times and record assignments accepted under $\varphi_1$ and $\varphi_2$, respectively.  Table \ref{tab:metrics} shows the sample means of accepted Mahalanobis distances under three assignment mechanisms: rerandomization with $\varphi_1$, rerandomization with $\varphi_2$, and complete randomization. The result justifies Corollary \ref{cor:epvr}, as $\varphi_1$ achieves the most balanced extent on each metric.
	\begin{table}
		\centering
		\caption{Sample Means of Accepted Mahalanobis Distances under Different Assignment Mechanisms.}
		\begin{tabular}{ccccccc}
			\hline
			Balance Criteria & $\mathbb{E}[D_1\mid\varphi]$ & $\mathbb{E}[D_2\mid\varphi]$ & $\mathbb{E}[D_3\mid\varphi]$ & $\mathbb{E}[D_4\mid\varphi]$ & $\mathbb{E}[D_5\mid\varphi]$ & $\mathbb{E}[D_6\mid\varphi]$ \\
			\hline
			$\varphi_1$ & 2.16 & 85.34 & 18.86 & 0.99 & 38.13 & 7.78 \\
			$\varphi_2$ & 2.65 & 88.60 & 23.25 & 1.04 & 45.57 & 12.46 \\
			$\backslash$ & 8.41 & 110.51 & 33.89 & 4.34 & 55.90 & 15.36 \\
			\hline
		\end{tabular}
		\label{tab:metrics}
	\end{table}
	
\begin{figure}[h]
		\centering
		\includegraphics[width=0.8\textwidth]{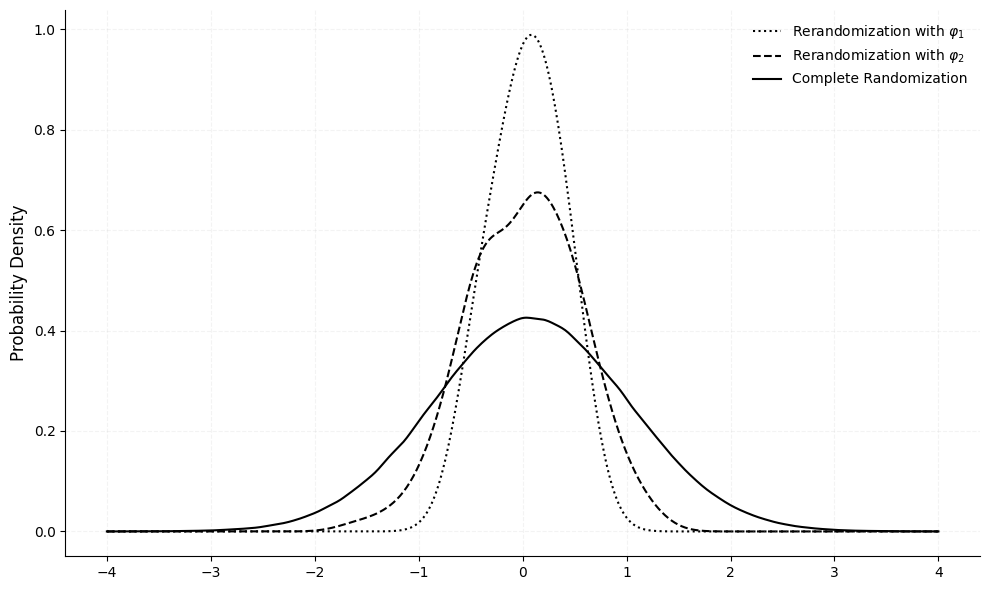}
		\caption{Scaled Factorial Contrasts of the Age Covariate Means under Different Assignment Mechanisms.}
		\label{fig:density}
	\end{figure}

	Figure \ref{fig:density} displays the distribution of the first scaled factorial contrast of the Age covariate means under the three assignment mechanisms. The distribution under $\varphi_1$ is the most peaked, indicating the most balanced extent, followed by $\varphi_2$ and then complete randomization.
	\section{Discussion and Conclusions}
	\label{sec:d&c}
	We evaluate balance criteria assuming fixed asymptotic acceptance probability. Some authors \cite[e.g.][]{wang2022rerandomization,harshaw2024balancing} consider diminishing acceptance probability. Although covariate imbalance would disappear asymptotically in the latter framework, it exists in finite sample when perfect balance is not achieved. \cite{harshaw2024balancing} find that ignoring the finite sample covariate imbalance can invalidate the subsequent analysis for causal effects. We keep the truncated sum of normal distributions to reflect covariate imbalance in finite sample.
	\cite{wang2022rerandomization} also consider diverging number of covariates. In practice, this adds to the total cost of conducting randomized experiments. A general framework to study the trade off between efficiency gain and the cost of collecting covariates is our future research topic. 
	
	Our decision-theoretic framework applies to general rerandomization procedures. For example, \cite{banerjee2020theory} formalizes another popular balance criterion that picks the best out of a pre-specified number of randomizations to maximize a utility function. Applying Theorem \ref{complete_class}, we find it inadmissible in terms of maximizing the conditional utility of accepted randomizations, as balance criteria in the complete class are deterministic almost surely besides at the decision boundary.
	
	In summary, conditional affine invariance and admissibility are two important principles in designing balance criteria for rerandomization. The former results in conditionally EPVR property for linear contrasts of covariate means, and the latter ensures that the asymptotic variance reduction cannot be dominated. Our theory helps researchers build balance criteria suitable for their experimental designs, and avoid inadmissible balance criteria that have already gained popularity. In particular, we recommend ReWM that follows the two principles, which applies to experimental designs involving tiers of covariates, stratification, and multiple treatment arms.
	\bibliographystyle{agsm}
	
	\bibliography{document}
	\newpage
	\appendix
	\section*{Supplementary Material} % Main title
	\setcounter{section}{0} % Reset section counter for S1, S2, etc.
	\renewcommand{\thesection}{S\arabic{section}} % Label sections as S1, S2, ...
	\counterwithout{lemma}{section} % Detach lemma counter from section
	\setcounter{lemma}{0}           % Reset lemma counter
	\renewcommand{\thelemma}{S\arabic{lemma}} % Lemmas labeled S1, S2,...
	\section{More Technical Details}\label{sec:detail}
	
	Extending Lemma A4 of \cite{li2018asymptotic}, we first state a lemma characterizing the centrality of the limiting distribution \eqref{pos_rewm} with respect to $\boldsymbol{\rho}$ for fixed $a$.
	\begin{lemma}
		\label{concentration}
		For any fixed \( a > 0 \), \( c \geq 0 \), \(\pi_m\geq0\) and \( m \in \{1, \ldots, M\} \), the probability
		\begin{equation}
			\label{tail_prob}
			\Pr\left(\sum_{m=1}^M \rho_m\epsilon^{(m)}_1+\sqrt{1 - \sum_{m=1}^M \rho_m^2}\cdot \epsilon_0\geq c \mid \left\{\sum_{m=1}^M \pi_m \Vert\boldsymbol{\epsilon}^{(m)}\Vert^2 \leq a\right\} \right)
		\end{equation}
		is non-increasing in each \( \rho_m \) for $\sum_{m=1}^M \rho_m^2\leq1$.
	\end{lemma}
	
	\begin{proof}
		We first prove the case where each $\boldsymbol{\epsilon}^{(m)}$ is scalar, so that \eqref{tail_prob} reduces to
		\begin{equation}
			\label{tail_prob_1}
			\Pr\left(\rho_m\epsilon_{m}+ \sqrt{1 - \sum_{m=1}^M \rho_m^2} \cdot \epsilon_0\geq c\mid\sum_{m=1}^{M}\pi_m\epsilon_{m}^2\leq a\right).
		\end{equation}
		Let \(\phi_0(\cdot)\) and \(\phi_{M,a}(\cdot)\) denote the density functions of $N(0,1)$ and the truncated normal distribution
		\(
		\left(\epsilon_{1},\ldots,\epsilon_{M}\right)\mid\{\sum_{m=1}^{M}\pi_m\epsilon_m^2 \leq a\}.
		\)
		The probability \eqref{tail_prob_1} can be expressed as
		\begin{equation*}
			P=\int_{\mathbb{R}^m} \Pr\left(\epsilon_0\geq\frac{c -\sum_{m=1}^{M}\rho_m\xi_m}{\sqrt{1 - \sum_{m=1}^{M}\rho_m^2} }\right)\phi_{M,a}(\boldsymbol{\xi})d\boldsymbol{\xi}
		\end{equation*}
		for $\boldsymbol{\xi}=(\xi_1,\ldots,\xi_M)$. Taking derivative with respect to $\rho_1$, we have
		\begin{equation*}
			\frac{\partial P}{\partial\rho_1}=\kappa\int I(R_{M,a})\exp\left(-\frac{1}{2}\left(\frac{\left(\xi_1'-c'\right)^2}{1-\sum_{m=1}^{M}\rho_m^2}+\sum_{m=2}^{M}\xi_m^{\prime2}\right)\right)(\xi_1'-c')d\boldsymbol{\xi}'
		\end{equation*}
		with $R_{M,a}$ determined by
		\begin{equation*}
			\pi_1\left(\frac{\xi_1'}{\sqrt{1-\sum_{m=2}^{M}\rho_m^2}}-\frac{\sum_{m=2}^{M}\rho_m\xi_m'}{1-\sum_{m=2}^{M}\rho_m^2}\right)^2+\sum_{m=2}^{M}\pi_{m}\xi_m^{\prime2}\leq a,
		\end{equation*}
		where
		\begin{equation*}
			\begin{aligned}
				\kappa&=\Pr\left(\sum_{m=1}^{M}\pi_m\xi_m^2 \leq a\right)^{-1}\exp\left(\frac{-\left(c-\sum_{m=2}^{M}\rho_m\xi_m\right)^2}{2\left(1-\sum_{m=2}^{M}\rho_m^2\right)}\right)\frac{\left(1-\sum_{m=2}^{M}\rho_m^2\right)^{1/2}}{\left(1-\sum_{m=1}^{M}\rho_m^2\right)^{3/2}}\left|\operatorname{det}\left(\frac{\partial\boldsymbol{\xi}}{\partial\boldsymbol{\xi}'}\right)\right|\geq0,\\
				\boldsymbol{\xi}^{\prime}&=(\xi_1^{\prime},\xi_2^{\prime}\ldots,\xi_M^{\prime})=\left(\sqrt{1-\sum_{m=2}^{M}\rho_m^2}\cdot \xi_1+\frac{\sum_{m=2}^{M}\rho_m\xi_m}{\sqrt{1-\sum_{m=2}^{M}\rho_m^2}},\xi_2,\ldots,\xi_M\right),\\
				c'&=\frac{\rho_1c}{\sqrt{1-\sum_{m=2}^{M}\rho_m^2}}.
			\end{aligned}
		\end{equation*}
		For any fixed \(\left(\xi_2',\ldots,\xi_M'\right)\), the range for $\xi_1'$ within the region $R_{M,a}$ can be derived as
		\begin{equation*}
			R(\xi_1'\mid\xi_2',\ldots,\xi_M')=\frac{\sum_{m=2}^{M}\rho_m\xi_m'}{\sqrt{1-\sum_{m=2}^{M}\rho_m^2}}+\left[-\sqrt{a-\sum_{m=2}^{M}\pi_{m}\xi_m^{\prime2}},\sqrt{a-\sum_{m=2}^{M}\pi_{m}\xi_m^{\prime2}}\right].
		\end{equation*}
		We consider the sum of integrals
		\begin{equation}
			\label{cond_int}
			\int_{R(\xi_1'\mid\xi_2',\ldots,\xi_M')}+\int_{R(\xi_1'\mid-\xi_2',\ldots,-\xi_M')}\exp\left(-\frac{1}{2}\left(\frac{\left(\xi_1'-c'\right)^2}{1-\sum_{m=1}^{M}\rho_m^2}\right)\right)(\xi_1'-c')d\xi_1'.
		\end{equation}
		Because the integrand is centrally symmetric around $(c',0)$ for $c'\geq0$, we know that \eqref{cond_int} is non-positive.
		By Fubini's Theorem and symmetry of $R_{M,a}$, we have
		\begin{equation*}
			\begin{aligned}
				2\frac{\partial P}{\partial\rho_1}&=\kappa\int\exp\left(-\sum_{m=2}^{M}\frac{\xi_m^{\prime2}}{2}\right)\int_{R(\xi_1'\mid\xi_2',\ldots,\xi_M')} \exp\left(-\frac{1}{2}\left(\frac{\left(\xi_1'-c'\right)^2}{1-\sum_{m=1}^{M}\rho_m^2}\right)\right)(\xi_1'-c')d\boldsymbol{\xi}'\\
				&\quad+\kappa\int\exp\left(-\sum_{m=2}^{M}\frac{\xi_m^{\prime2}}{2}\right)\int_{R(\xi_1'\mid-\xi_2',\ldots,-\xi_M')} \exp\left(-\frac{1}{2}\left(\frac{\left(\xi_1'-c'\right)^2}{1-\sum_{m=1}^{M}\rho_m^2}\right)\right)(\xi_1'-c')d\boldsymbol{\xi}'\\
				&\leq0.
			\end{aligned}
		\end{equation*}
		Consequently, the probability \eqref{tail_prob_1} is non-decreasing in $\rho_1$ for $\sum_{m=1}^M \rho_m^2\leq1$, which is also true for the remaining correlation coefficients by applying the same arguments.
		
		In general, for $1\leq m\leq M$, let $\boldsymbol{\epsilon}_{-1}^{(m)}$ denote the subvector that removes the first element of $\boldsymbol{\epsilon}_{m}$. Then \eqref{tail_prob} can be written as
		\begin{equation*}
			\mathbb{E}\left[\Pr\left(\rho_m\epsilon^{(m)}_1+ \sqrt{1 - \sum_{m=1}^M \rho_m^2} \cdot \epsilon_0\geq c\mid\sum_{m=1}^{M}\pi_m\epsilon^{(m)2}_1\leq a-\sum_{m=1}^{M}\Vert\boldsymbol{\epsilon}_{-1}^{(m)}\Vert^2\right)\right].
		\end{equation*}
		Condition on the remaining elements, the inner probability reduces to the scalar case where the monotonicity holds in each $\rho_m$ for $\sum_{m=1}^{M}\rho_m^2\leq1$. By iterated expectation, the conclusion generates to the multivariate case.
	\end{proof}
	
	Given $J$ strata, for $1\leq j\leq J$, let $s_{jq}$ denote the sample variance of the observed outcomes:
	\begin{equation*}
		s_{jq}=\frac{1}{n_{jq}-1}\sum_{\boldsymbol{x}_i^{(s)}=\boldsymbol{c}(j),W_{iq}=1}\left(y_i-\overline{y}_{q}^{(j)}\right)^2,\quad0\leq q\leq Q-1,
	\end{equation*}
	where $\overline{y}_{q}^{(j)}$ denotes the sample mean of $y_i$ given $\boldsymbol{x}_i^{(s)}=\boldsymbol{c}(j)$ and $W_{iq}=1$ for $W_{i0}=1-\sum_{q=1}^{Q-1}W_{iq}$, and $n_{jq}$ denote the number of units in the $j$-th stratum under the $q$-th treatment.
	Let $\boldsymbol{s}_{\boldsymbol{x},jq}$ denote the sample covariance between the observed outcomes and the remaining covariates besides stratum indicators:
	\begin{equation*}
		\boldsymbol{s}_{\boldsymbol{x},jq}=\frac{1}{n_{jq}-1}\sum_{\boldsymbol{x}_i^{(s)}=\boldsymbol{c}(j),W_{iq}=1}\left(y_i-\overline{y}_{q}^{(j)}\right)\left(\boldsymbol{x}_i^{(r)}-\overline{\boldsymbol{x}}^{(jr)}\right),\quad0\leq q\leq Q-1.
	\end{equation*}
	Let $\boldsymbol{S}_{\boldsymbol{x},j}$ denote the sample covariance of the remaining covariates:
	\begin{equation*}
		\boldsymbol{S}_{\boldsymbol{x},j}=\frac{1}{n_{j}-1}\sum_{\boldsymbol{x}_i^{(s)}=\boldsymbol{c}(j)}\left(\boldsymbol{x}_i^{(r)}-\overline{\boldsymbol{x}}^{(jr)}\right)^{\top}\left(\boldsymbol{x}_i^{(r)}-\overline{\boldsymbol{x}}^{(jr)}\right),\quad0\leq q\leq Q-1.
	\end{equation*}
	Define
	\begin{equation*}
		\widehat{\boldsymbol{\Sigma}}_{\tau}=2^{-2(K-1)}\sum_{j=1}^{J}\sum_{q=0}^{Q-1}\frac{n_{j}^2}{nn_{jq}}\boldsymbol{d}_q\boldsymbol{d}_q^{\top}\otimes s_{jq},\ \widehat{\boldsymbol{\Sigma}}_{\tau,\boldsymbol{x}}=2^{-2(K-1)}\sum_{j=1}^{J}\sum_{q=0}^{Q-1}\frac{n_{j}^2}{nn_{jq}}\boldsymbol{d}_q\boldsymbol{d}_q^{\top}\otimes\boldsymbol{s}_{\boldsymbol{x},jq}
	\end{equation*}
	and
	\begin{equation*}
		\boldsymbol{\Sigma}_{\boldsymbol{x}}=2^{-2(K-1)}\sum_{j=1}^{J}\sum_{q=0}^{Q-1}\frac{n_{j}^2}{nn_{jq}}\boldsymbol{d}_q\boldsymbol{d}_q^{\top}\otimes\boldsymbol{S}_{\boldsymbol{x},j}.
	\end{equation*}
	where $\otimes$ denotes the Kronecker product between two matrices, and $\boldsymbol{d}_q$ ($0\leq q\leq Q-1$) denotes the $(q+1)$-th column of $\left(\boldsymbol{g}_1^{\top},\ldots,\boldsymbol{g}_F^{\top}\right)^{\top}$. \cite{liu2024randomization} show that, under certain regularity conditions,
	\begin{equation}
		\label{estimator}
		\begin{aligned}
			&\widehat{\boldsymbol{\Sigma}}_{\tau}=\operatorname{cov}\left(\sqrt{n}\widehat{\boldsymbol{\tau}}\right)+\Delta+o_p(1)\text{ for }\Delta\geq0,\\
			&\widehat{\boldsymbol{\Sigma}}_{\tau,\boldsymbol{x}}=\operatorname{cov}\left(\sqrt{n}\widehat{\boldsymbol{\tau}},\sqrt{n}\widehat{\boldsymbol{\tau}}_{\boldsymbol{x}}\right)+o_p(1),\\
			\text{and }&\boldsymbol{\Sigma}_{\boldsymbol{x}}=\operatorname{cov}\left(\sqrt{n}\widehat{\boldsymbol{\tau}}_{\boldsymbol{x}}\right)
		\end{aligned}
	\end{equation}
	under stratified completely randomized $2^K$ factorial design. The following lemma guarantees that \eqref{estimator} still holds under ReWM. Therefore, we can use $n\widehat{v}=\boldsymbol{\lambda}\widehat{\boldsymbol{\Sigma}}_{\tau}\boldsymbol{\lambda}^{\top}$ as an asymptotically conservative estimator for $\operatorname{var}(\sqrt{n}\widehat{\tau})$ under ReWM, where $\boldsymbol{\lambda}=\left(\lambda_1,\ldots,\lambda_F\right)$ defines $\widehat{\tau}=\sum_{f=1}^{F}\lambda_f\widehat{\tau}_f$.
	
	\begin{lemma}
		\label{dominated_converge_in_p}
		For any sequence of random variables, $\{RV_n\}_{n=1}^{\infty}$, that is $o_p(n^k)$ for some real number $k$ under CRE, it is also $o_p(n^k)$ under ReWM as $n\rightarrow\infty$.
	\end{lemma}
	\begin{proof}
		The proof is identical to Lemma S1 of \cite{zhong2024two}, so we omit it.
	\end{proof}
	
	For $1\leq m\leq M$, denote by $\boldsymbol{z}^{(1:m)}=\left(\boldsymbol{z}^{(1)},\ldots,\boldsymbol{z}^{(m)}\right)$. After reordering the scaled linear contrasts of covariates, we can write $\boldsymbol{z}^{(1:M)}=\sqrt{n}\widehat{\boldsymbol{\tau}}_{\boldsymbol{x}}\boldsymbol{P}$ with $\operatorname{cov}\left(\boldsymbol{z}^{(1:M)}\right)=\boldsymbol{P}^{\top}\boldsymbol{\Sigma}_{\boldsymbol{x}}\boldsymbol{P}$, where $\boldsymbol{P}$ is a permutation matrix. Then $\operatorname{cov}\left(\boldsymbol{z}^{(1:m-1)}\right)$ is the submatrix of $\boldsymbol{P}^{\top}\boldsymbol{\Sigma}_{\boldsymbol{x}}\boldsymbol{P}$ with rows and columns indexed from $1$ to $\sum_{\ell=1}^{m-1}k_{\ell}$; $\operatorname{cov}\left(\boldsymbol{z}^{(m)},\boldsymbol{z}^{(1:m-1)}\right)$ is the submatrix with rows indexed from $\sum_{\ell=1}^{m-1}k_{\ell}+1$ to $\sum_{\ell=1}^{m}k_{\ell}$, and columns indexed from $1$ to $\sum_{\ell=1}^{m}k_{\ell}$; and $\operatorname{cov}\left(\boldsymbol{z}^{(m)}\right)$ is the submatrix with rows and columns indexed from $\sum_{\ell=1}^{m-1}k_{\ell}+1$ to $\sum_{\ell=1}^{m}k_{\ell}$. For $1<m\leq M$, the block-wise Gram-Schmidt orthogonalization is given by
	\begin{equation*}
		\widetilde{\boldsymbol{z}}^{(m)\top}= \boldsymbol{z}^{(m)\top} - \operatorname{cov}\left(\boldsymbol{z}^{(m)},\boldsymbol{z}^{(1:m-1)}\right)\operatorname{cov}\left(\boldsymbol{z}^{(1:m-1)}\right)^{-1}\boldsymbol{z}^{(1:m-1)\top},
	\end{equation*}
	which satisfies $\operatorname{cov}\left(\widetilde{\boldsymbol{z}}^{(m)}\right)=\operatorname{cov}\left(\boldsymbol{z}^{(m)}\right)-\operatorname{cov}\left(\boldsymbol{z}^{(m)},\boldsymbol{z}^{(1:m-1)}\right)\operatorname{cov}\left(\boldsymbol{z}^{(1:m-1)}\right)^{-1}\operatorname{cov}\left(\boldsymbol{z}^{(m)},\boldsymbol{z}^{(1:m-1)}\right)^{\top}$, and $\operatorname{cov}\left(\widetilde{\boldsymbol{z}}^{(m)},\boldsymbol{z}^{(1:m-1)}\right)=\boldsymbol{0}$. It follows that $\widetilde{\boldsymbol{z}}^{(1)},\ldots,\widetilde{\boldsymbol{z}}^{(M)}$ are mutually uncorrelated with
	\begin{equation*}
		\sum_{\ell=1}^{m}D_{\ell}=\sum_{\ell=1}^{m}\widetilde{\boldsymbol{z}}^{(\ell)}\operatorname{cov}\left(\widetilde{\boldsymbol{z}}^{(\ell)}\right)^{-1}\widetilde{\boldsymbol{z}}^{(\ell)\top}=\widetilde{\boldsymbol{z}}^{(1:m)}\operatorname{cov}\left(\widetilde{\boldsymbol{z}}^{(1:m)}\right)^{-1}\widetilde{\boldsymbol{z}}^{(1:m)\top},
	\end{equation*}
	where $\widetilde{\boldsymbol{z}}^{(1:m)}=\left(\widetilde{\boldsymbol{z}}^{(1)},\ldots,\widetilde{\boldsymbol{z}}^{(m)}\right)$, $(1< m\leq M)$. Since each $\widetilde{\boldsymbol{z}}^{(1:m)}$ is an affine transformation of $\boldsymbol{z}^{(1:m)}$, we also have
	\begin{equation*}
		\sum_{\ell=1}^{m}D_{\ell}=\boldsymbol{z}^{(1:m)}\operatorname{cov}\left(\boldsymbol{z}^{(1:m)}\right)^{-1}\boldsymbol{z}^{(1:m)\top}.
	\end{equation*}
	Therefore, we can calculate $D_1=\boldsymbol{z}^{(1)}\operatorname{cov}\left(\boldsymbol{z}^{(1)}\right)^{-1}\boldsymbol{z}^{(1)\top}$ and
	\begin{equation*}
		D_m=\boldsymbol{z}^{(1:m)}\operatorname{cov}\left(\boldsymbol{z}^{(1:m)}\right)^{-1}\boldsymbol{z}^{(1:m)\top}-\boldsymbol{z}^{(1:m-1)}\operatorname{cov}\left(\boldsymbol{z}^{(1:m-1)}\right)^{-1}\boldsymbol{z}^{(1:m-1)\top},\quad1<m\leq M.
	\end{equation*}
	Moreover, the squared multiple correlation coefficient between $v^{-1/2}\left(\widehat{\tau}- \tau\right)$ and $\widetilde{\boldsymbol{z}}^{(1:m)}$ is
	\begin{equation*}
		\begin{aligned}
			\sum_{\ell=1}^{m}\rho_\ell^2&=\sum_{\ell=1}^{m}\frac{\operatorname{cov}\left(\sqrt{n}\widehat{\tau},\widetilde{\boldsymbol{z}}^{(\ell)}\right)\operatorname{cov}\left(\widetilde{\boldsymbol{z}}^{(\ell)}\right)^{-1}\operatorname{cov}\left(\sqrt{n}\widehat{\tau},\widetilde{\boldsymbol{z}}^{(\ell)}\right)^{\top}}{\operatorname{var}\left(\sqrt{n}\widehat{\tau}\right)}\\&=\frac{\operatorname{cov}\left(\sqrt{n}\widehat{\tau},\widetilde{\boldsymbol{z}}^{(1:m)}\right)\operatorname{cov}\left(\widetilde{\boldsymbol{z}}^{(1:m)}\right)^{-1}\operatorname{cov}\left(\sqrt{n}\widehat{\tau},\widetilde{\boldsymbol{z}}^{(1:m)}\right)^{\top}}{\operatorname{var}\left(\sqrt{n}\widehat{\tau}\right)}\\
			&=\frac{\operatorname{cov}\left(\sqrt{n}\widehat{\tau},\boldsymbol{z}^{(1:m)}\right)\operatorname{cov}\left(\boldsymbol{z}^{(1:m)}\right)^{-1}\operatorname{cov}\left(\sqrt{n}\widehat{\tau},\boldsymbol{z}^{(1:m)}\right)^{\top}}{\operatorname{var}\left(\sqrt{n}\widehat{\tau}\right)}.
		\end{aligned}
	\end{equation*}
	Using \eqref{estimator} and Lemma \ref{dominated_converge_in_p}, $\operatorname{cov}\left(\sqrt{n}\widehat{\tau},\sqrt{n}\widehat{\boldsymbol{\tau}}_{\boldsymbol{x}}\boldsymbol{P} \right)$ can be consistently estimated by $\boldsymbol{\lambda}\widehat{\boldsymbol{\Sigma}}_{\tau,\boldsymbol{x}}\boldsymbol{P}$ under ReWM. We then obtain an estimator for $\sum_{\ell=1}^{m}\rho_\ell^2$ as
	\begin{equation*}
		\sum_{\ell=1}^{m}\widehat{\rho}_\ell^2=\frac{\widehat{\operatorname{cov}}\left(\sqrt{n}\widehat{\tau},\boldsymbol{z}^{(1:m)}\right)\operatorname{cov}\left(\boldsymbol{z}^{(1:m)}\right)^{-1}\widehat{\operatorname{cov}}\left(\sqrt{n}\widehat{\tau},\boldsymbol{z}^{(1:m)}\right)^{\top}}{n\widehat{v}},\quad1\leq m\leq M,
	\end{equation*}
	where $\widehat{\operatorname{cov}}\left(\sqrt{n}\widehat{\tau},\boldsymbol{z}^{(1:m)}\right)$ is the sub-vector of $\boldsymbol{\lambda}\widehat{\boldsymbol{\Sigma}}_{\tau,\boldsymbol{x}}\boldsymbol{P}$ indexed from $1$ to $\sum_{\ell=1}^{m}$.
	The continuous mapping theorem and \eqref{estimator} imply that $\sum_{\ell=1}^{m}\widehat{\rho}_\ell^2=\sum_{\ell=1}^{m}\rho_\ell^2-\Delta'+o_p(1)$ for $\Delta'\geq0$ under ReWM.

	Let $L_{\alpha/2}(\widehat{\boldsymbol{\rho}},a)$ denote the $(1-\alpha/2)$-th quantile of
	\begin{equation}
		\label{pos_rewm_est}
		\sum_{m=1}^M \widehat{\rho}_m\epsilon^{(m)}_1 \mid \left\{\sum_{m=1}^M \pi_m \Vert\boldsymbol{\epsilon}^{(m)}\Vert^2 \leq a\right\}+\sqrt{1 - \sum_{m=1}^M \widehat{\rho}_m^2}\cdot \epsilon_0,
	\end{equation}
	which is asymptotically no smaller than $L_{\alpha/2}(\boldsymbol{\rho},a)$ for \eqref{pos_rewm} according to Lemma \ref{concentration}. Therefore, an asymptotically conservative \(1-\alpha\) confidence interval for $\tau$ under ReWM can be given by
	\[
	\left( \widehat{\tau} - \sqrt{\widehat{v}} \, L_{\alpha/2}(\widehat{\boldsymbol{\rho}},a), \widehat{\tau} + \sqrt{\widehat{v}} L_{\alpha/2}(\widehat{\boldsymbol{\rho}},a) \right).
	\]
	In practice, we can simulate \eqref{pos_rewm_est} by drawing random samples from independent normal distributions $\{\boldsymbol{\epsilon}^{(m)}\}_{m=1}^{M}$ and $\epsilon_0$, and determine $L_{\alpha/2}(\widehat{\boldsymbol{\rho}},a)$ to be the $(1-\alpha/2)$-th quantile of the simulated distribution of \eqref{pos_rewm_est}.
	
	\section{Proofs to Theorems and Propositions}\label{sec:proof}
	\begin{proof}[Proof of Theorem \ref{complete_class}]
		Let $\sigma$ denote the underlying probability measure.
		
		\noindent$(\mathrm{\romannumeral1})$ We first prove the case of $M=1$. Let $\varphi'$ be another criterion in $\mathcal{C}$ with $\mathbb{E}[\varphi']\geq p$. Denote by $S^{+}$ and $S^{-}$ the sets in $\mathcal{W}$ where $\varphi-\varphi'>0$ and $<0$ respectively. If $\omega$ in $S^{+}$, $\varphi(\omega)$ must be positive and $a\geq{}\pi_1\zeta_1(\omega)$. In a similar way $a\leq{}\pi_1\zeta_1(\omega)$ for all $\omega$ in $S^{-}$. Hence,
		\begin{equation*}
			\int\left(a-\pi_1\zeta_1\right)\left(\varphi-\varphi'\right) d\sigma=\int_{S^{+}\cup S^{-}}\left(a-\pi_1\zeta_1\right)\left(\varphi-\varphi'\right) d\sigma \geq 0.
		\end{equation*}
		It follows that $\int{}\zeta_1\varphi d\sigma-\int{}\zeta_1\varphi' d\sigma\leq a/\pi_1\left(\int\varphi{}d\sigma-\int\varphi'd\sigma\right)$ and that
		\begin{equation*}
			\begin{aligned}
				\mathbb{E}[\zeta_1\mid\varphi]-\mathbb{E}[\zeta_1\mid\varphi']&=\frac{\int{}\zeta_1\varphi{}d\sigma}{\int\varphi{}d\sigma}-\frac{\int{}\zeta_1\varphi'd\sigma}{\int\varphi'd\sigma}\\
				&=\frac{\int{}\zeta_1\varphi{}d\sigma+\mathbb{E}[\zeta_1\mid\varphi]\left(\int\varphi'd\sigma-\int\varphi{}d\sigma\right)}{\int\varphi'd\sigma}-\frac{\int{}\zeta_1\varphi'd\sigma}{\int\varphi'd\sigma}\\
				&\leq\frac{\left(\mathbb{E}[\zeta_1\mid\varphi]-a/\pi_1\right)\left(\int\varphi'd\sigma-\int\varphi{}d\sigma\right)}{\int\varphi'd\sigma}\\
				&\leq0,
			\end{aligned}
		\end{equation*}
		where the last inequality is because $a\geq\pi_1\zeta_1$ on $\varphi>0$ and $\mathbb{E}[\varphi']\geq\mathbb{E}[\varphi]$.
		
		We then prove the $M>1$ case by contradiction. Denote by $\boldsymbol{\pi}\cdot\boldsymbol{\zeta}=\sum_{m=1}^{M}\pi_m{}\zeta_m$. Suppose there exists $\varphi'\in\mathcal{C}$ that dominates $\varphi$. Summing these inequalities with respect to the positive vector $\boldsymbol{\pi}$, we have $\mathbb{E}\left[\boldsymbol{\pi}\cdot\boldsymbol{\zeta}\mid\varphi'\right]<\mathbb{E}\left[\boldsymbol{\pi}\cdot\boldsymbol{\zeta}\mid\varphi\right]$. However, as shown in the $M=1$ case, $\mathbb{E}\left[\boldsymbol{\pi}\cdot\boldsymbol{\zeta}\mid\varphi'\right]\geq\mathbb{E}\left[\boldsymbol{\pi}\cdot\boldsymbol{\zeta}\mid\varphi\right]$. This contradiction means that $\varphi'$ does not exist and $\varphi$ is an admissible criterion.
		
		\noindent$(\mathrm{\romannumeral2})$
		For all $\psi$ in $\mathcal{C}$, let $U$ be the set of points in $\mathbb{R}^{M+1}$:
		\begin{equation}
			\left\{\left(\int \zeta_1\psi d\sigma, \ldots, \int \zeta_M\psi d\sigma,  \int1-\psi d\sigma\right):\psi\in\mathcal{C}\right\}.
		\end{equation}
		Note that $U$ is convex because it is the image of a linear transformation on the convex set $\mathcal{C}$. Suppose an admissible criterion $\varphi\in\mathcal{C}$ is mapped to some $u=\left(u_1,\ldots,u_{M+1}\right)\in{}U$. Let $V$ be the set of all points whose first $M$ coordinates are strictly lower than those of $u$ for fixed $u_{M+1}$. We state that $U\cap{}V=\emptyset$. Otherwise, we can find a point in $U$ mapped from $\varphi'$ such that $\mathbb{E}[\varphi']=u_{M+1}$ and $\int \zeta_m\varphi'd\sigma<\int \zeta_m\varphi d\sigma$, $m=1,2,\ldots,M$, which means $\varphi$ is no longer admissible, a contradiction.
		
		By the separation theorem there exists a hyperplane $\Pi$ that separates two disjoint convex sets $U$ and $V$. An immediate observation is $u\in\Pi$ because $u\in{}U\cap\overline{V}$. Let the equation of $\Pi$ evaluated at $u$ be $\pi_0+\sum_{m=1}^{M}\pi_m{}u_m+au_{M+1}=0$. According to the analytical form of the separation theorem,
		\begin{align}
			\label{sep}
			&\sum_{m=1}^{M}\pi_m\left(v_m-u_m\right)\leq0,\quad\forall(v_1,\ldots,v_m, u_{M+1})\in V,\\
			\label{sep:phi}
			&\int\left(\boldsymbol{\pi}\cdot\boldsymbol{\zeta}-a\right)\psi d\sigma\geq\int\left(\boldsymbol{\pi}\cdot\boldsymbol{\zeta}-a\right)\psi d\sigma,\quad\forall\psi\in\mathcal{C}.
		\end{align}
		It can be directly proved that $\pi_m\geq0$ ($m=1,2,\ldots,M$) by letting each coordinate tend to $-\infty$ with the rest fixed in (\ref{sep}).
		
		Consider $\varphi^*\in\mathcal{C}$ such that $\varphi^*(\omega)=1$ when $\boldsymbol{\pi}\cdot\boldsymbol{\zeta}(\omega)\leq a$ and $\varphi^*(\omega)=0$ when $\boldsymbol{\pi}\cdot\boldsymbol{\zeta}(\omega)>a$. Then $\varphi^*$ attains the lower bound of \eqref{sep:phi}. Let $S$ be the intersection of the set, on which $\varphi^*$ and $\varphi$ differs, with $\{\omega\in\mathcal{W}\mid{}\boldsymbol{\pi}\cdot\boldsymbol{\zeta}(\omega)\neq{}a\}$. Since $\left(\boldsymbol{\pi}\cdot\boldsymbol{\zeta}-a\right)\left(\varphi^*-\varphi\right)>0$ on $S$ and equals $0$ on $\mathcal{W}\backslash{}S$, we have
		\begin{equation*}
			\begin{aligned}
				0&\leq\int_{S}\left(\boldsymbol{\pi}\cdot\boldsymbol{\zeta}-a\right)\left(\varphi^*-\varphi\right)d\sigma\\
				&=\int\left(\boldsymbol{\pi}\cdot\boldsymbol{\zeta}-a\right)\left(\varphi^*-\varphi\right)d\sigma=0
			\end{aligned}
		\end{equation*}
		Thus, $\int_{S}\left(\boldsymbol{\pi}\cdot\boldsymbol{\zeta}-a\right)\left(\varphi^*-\varphi\right)d\sigma=0$ which implies that $\Pr(S)=0$ as was to be proved.
		
		\noindent$(\mathrm{\romannumeral3})$
		For all $\psi\in\mathcal{C}$, let $U_{\text{ext}}$ be the set of points in $\mathbb{R}^{2M+2}$:
		\begin{align*}
			\Bigg\{\Bigg(&\int \zeta_1\psi d\sigma, \ldots, \int \zeta_M\psi d\sigma, \int\psi d\sigma, \int \zeta_1(1-\psi) d\sigma, \ldots,\\
			&\int \zeta_M(1-\psi) d\sigma, \int1-\psi d\sigma\Bigg):\psi\in\mathcal{C}\Bigg\}
		\end{align*} 
		Applying Theorem 1 of \cite{dvoretzky1951relations}, $U_{\text{ext}}$ is convex and compact. Recall that $U$ is defined as $$U=\left\{\left(\int \zeta_1\psi d\sigma, \ldots, \int \zeta_M\psi d\sigma,  \int1-\psi d\sigma\right):\psi\in\mathcal{C}\right\}.$$ Because $U$ can be obtained by coordinate projection over $U_{\text{ext}}$, we know that $U$ is still convex and compact.
		
		For any $\varphi\in\mathcal{C}\backslash\mathcal{C}^*$, by (ii), we know that $\varphi$ is inadmissible and can be dominated by some $\varphi'\in\mathcal{C}$. Consider a family of convex sets $V(d)=\{(v_1,\ldots,v_{M+1})\in\mathbb{R}^{M+1}\mid v_1\leq u_1'-d, \ldots, v_M\leq u_M'-d, v_{M+1}=u_{M+1}'\}$, where
		\begin{equation*}
			u'=(u_1',\ldots,u_M',u_{M+1}')=\left(\int \zeta_1\varphi' d\sigma,\ldots,\int \zeta_M\varphi' d\sigma, \int1-\varphi' d\sigma\right).
		\end{equation*}
		Define $d_0=\inf\left\{d\geq0:V(d)\cap U=\emptyset\right\}$. Because $U$ is bounded from below in each coordinate, we know that $q_0<\infty$.
		
		We state $V(d_0)\cap U\neq\emptyset$. If, on the contrary, $V(d_0)\cap U=\emptyset$, because $V(d_0)$ is convex and closed, there exists a hyperplane that strictly separates them. It indicates the existence of $d_1<d_0$ such that $V(d_1)\cap U=\emptyset$, which contradicts with the definition of $d_0$. Therefore, there exists $u^*\in V(d_0)\cap U$ for $d_0\geq0$, where
		\begin{equation*}
			u^*=\left(u^*_1,\ldots,u^*_M,u^*_{M+1}\right)=\left(\int \zeta_1\varphi^*d\sigma,\ldots,\int \zeta_M\varphi^*d\sigma, \int1-\varphi^*d\sigma\right)
		\end{equation*}
		satisfies $\mathbb{E}[\varphi^*]=\mathbb{E}[\varphi']$ and $\mathbb{E}[\zeta_m\mid\varphi^*]\leq\int \mathbb{E}[\zeta_m\mid\varphi']$ for $m=1,\ldots,M$. It follows that $\varphi^*$ also dominates $\varphi$. Applying the separation theorem to $V(d_0)$ and $U$, we repeat the proof of (ii) to obtain that
		\begin{equation*}
			\varphi^*(\omega)\stackrel{a.s.}{=}
			\begin{cases}
				1,& \sum_{m=1}^{M}\pi_m{}\zeta_m(\omega)<a, \\
				0,& \sum_{m=1}^{M}\pi_m{}\zeta_m(\omega)>a
			\end{cases}
		\end{equation*}
		for $\pi_1\geq0,\ldots,\pi_M\geq0, a\in\mathbb{R}$.
		Then $\varphi^*\in\mathcal{C}^*$.
	\end{proof}
	
	\begin{proof}[Proof of Proposition \ref{prop:cost}]
		Given a sample $\omega\in\mathcal{W}$ and an intersection criterion $\varphi$, $m(\omega,\varphi)$ is the number of steps when first $\varphi_m=0$ occurs (and equals $M$ if $\varphi_m=1$ at all steps). Therefore, the cost of making a decision on $\omega$ is $\sum_{m=1}^{m(\omega,\varphi)}h_m(\omega)=h_1(\omega)+\sum_{m=2}^{M}\left(h_m(\omega)\prod_{l=1}^{m-1}\varphi_{l}(\omega)\right)$. Then
		\begin{align*}
			\mathbb{E}\left[\sum_{m=1}^{m(\omega,\varphi)}h_m(\omega)\right]
			&=\mathbb{E}\left[h_1+\sum_{m=2}^{M}\left(h_m\prod_{l=1}^{m-1}\varphi_{l}\right)\right]\\
			&=\mathbb{E}[h_1]+\sum_{m=2}^{M}\mathbb{E}\left[h_m\prod_{l=1}^{m-1}\varphi_{l}\right].
		\end{align*}
		For $m\geq2$, since $h_m\geq0$ and $\prod_{l=1}^{m-1}\varphi_{l}\leq1$, we further know that $h_m\left(\prod_{l=1}^{m-1}\varphi_{l}-1\right)\leq0$. Taking expectation on both sides and summing over $m$, we have $\sum_{m=2}^{M}\mathbb{E}[h_m\prod_{l=1}^{m-1}\varphi_{l}]\leq\sum_{m=2}^{M}\mathbb{E}[h_m]$,
		which completes the proof.
	\end{proof}
	\begin{proof}[Proof of Corollary \ref{cor:epvr}]
		\noindent$(\romannumeral1)$ Equation \eqref{pos_rewm} can be proved in the same way as Proposition A1 of \cite{li2018asymptotic}. By Theorem \ref{shrink_cond}, we have
		\begin{equation*}
			\sum_{m=1}^M \rho_m\epsilon_1^{(m)} \mid \{\varphi=1\} + \sqrt{1 - \sum_{m=1}^M \rho_m^2} \cdot \epsilon_0\sim\sum_{m=1}^M \epsilon_{i_m}^{(m)} \mid \{\varphi=1\} + \sqrt{1 - \sum_{m=1}^M \rho_m^2} \cdot \epsilon_0
		\end{equation*}
		for $1\leq i_m\leq k_m$, $1\leq m\leq M$. Because each $\epsilon_{i_m}^{(m)}$ is symmetric around 0, we have
		\begin{equation*}
			\begin{aligned}
				\operatorname{var}_{\text{lim}}\left(v^{-1/2}\widehat{\tau} \mid \varphi = 1\right) &=\sum_{m=1}^M \frac{\rho_m^2}{k_m}\mathbb{E}_{\text{lim}}\left[\sum_{1\leq i_m\leq k_m}\left(\epsilon_{i_m}^{(m)}\right)^2 \mid \varphi=1\right] + \left(1 - \sum_{m=1}^M \rho_m^2\right)\\
				&=\sum_{m=1}^M \frac{\rho_m^2}{k_m}\mathbb{E}_{\text{lim}}[D_m \mid \varphi=1] + \left(1 - \sum_{m=1}^M \rho_m^2\right).
			\end{aligned}
		\end{equation*}
		\noindent$(\romannumeral2)$ The completeness of \(\mathcal{C}_M^*\) follows from Theorem \ref{complete_class}. With positive weights $\pi_1,\ldots,\pi_M$, we can also apply Theorem \ref{complete_class} to have
		\begin{equation*}
			\varphi^*=I\left( \sum_{m=1}^M \pi_m D_m \leq a \right)
		\end{equation*}
		is admissible for minimizing \(\{\mathbb{E}_{\text{lim}}\left[D_{m} \mid \varphi=1\right]\}_{m=1}^{M}\), provided that $\lim_{n\rightarrow\infty}\Pr(\varphi^*=1)=p$. It remains to generalize the conclusion to non-negative weights. If all weights are $0$, the acceptance probability is either $1$ or $0$, which are both trivial cases. Suppose that the positive weights are $\pi_{j_1},\ldots,\pi_{j_L}$ for $1\leq L< M$. Applying Theorem \ref{complete_class}, for minimizing \(\mathbb{E}_{\text{lim}}\left[\sum_{\ell=1}^{L}\pi_{j_\ell}D_{j_\ell} \mid \varphi=1\right]\),
		\begin{equation*}
			\varphi^{**}=I\left( \sum_{\ell=1}^L \pi_{j_\ell} D_{j_\ell} \leq a \right)
		\end{equation*}
		is admissible, provided that $\lim_{n\rightarrow\infty}\Pr(\varphi^{**}=1)=p$. Moreover, a necessary condition for any balance criterion $\varphi\in\mathcal{C}_M$ being admissible for minimizing \(\mathbb{E}_{\text{lim}}\left[\sum_{\ell=1}^{L}\pi_{j_\ell}D_{j_\ell} \mid \varphi=1\right]\) is that $\varphi=\varphi^{**}$ almost surely. Therefore, if any balance criterion $\varphi'\in\mathcal{C}_M$ dominates $\varphi^{**}$ in minimizing \(\{\mathbb{E}_{\text{lim}}\left[D_{m} \mid \varphi=1\right]\}_{m=1}^{M}\), it must have no larger conditional expectation than \(\mathbb{E}_{\text{lim}}\left[\sum_{\ell=1}^{L}\pi_{j_\ell}D_{j_\ell} \mid \varphi=1\right]\), which leads to $\varphi'=\varphi^{**}$ almost surely, a contradiction.\\
		\noindent$(\romannumeral3)$ The acceptance region defined by any balance criterion in \(\mathcal{C}_M^*\) is a pyramid defined by
		\begin{equation*}
			\sum_{m=1}^M \pi_m D_m \leq a,\quad D_m\geq0\ (1\leq m\leq M),\quad\text{a.s.}.
		\end{equation*}
		However, the acceptance region defined by $\prod_{m=1}^M I(D_m \leq a_m)$ is a rectangular region, which differs from any balance criterion in \(\mathcal{C}_M^*\) by a non-zero probability measure.
	\end{proof}
	
\end{document}